\documentclass[%
 reprint,
superscriptaddress,
 amsmath,amssymb,
 longbibliography,
 aps,
 prl,
]{revtex4-2}

\usepackage{graphicx}
\usepackage{dcolumn}
\usepackage{bm}
\usepackage{multirow}
\usepackage{xcolor}
\usepackage{hyperref}
\usepackage{orcidlink}

\usepackage{color}
\definecolor{BLUE}{rgb}{0.0,0.0,1.0}

\usepackage{soul}
\soulregister\cite7
\soulregister\ref7

\begin{document}

\title{Pinning down electron correlations in RaF via spectroscopy of excited states {and high-accuracy relativistic quantum chemistry}}

\author{M. Athanasakis-Kaklamanakis\orcidlink{0000-0003-0336-5980}}
 \email{m.athkak@cern.ch}
\affiliation{Experimental Physics Department, CERN, CH-1211 Geneva 23, Switzerland}
\affiliation{KU Leuven, Instituut voor Kern- en Stralingsfysica, B-3001 Leuven, Belgium}
\affiliation{Blackett Laboratory, Centre for Cold Matter, Imperial College London, SW7 2AZ London, United Kingdom}

\author{S. G. Wilkins}
 \email{wilkinss@mit.edu}
\affiliation{Department of Physics, Massachusetts Institute of Technology, Cambridge, MA 02139, USA}
\affiliation{Laboratory for Nuclear Science, Massachusetts Institute of Technology, Cambridge, MA 02139, USA}

\author{L. V. Skripnikov\orcidlink{0000-0002-2062-684X}}
 \email{skripnikov\_lv@pnpi.nrcki.ru}
\affiliation{Affiliated with an institute covered by a cooperation agreement with CERN.}

\author{\'{A}. Koszor\'{u}s\orcidlink{0000-0001-7959-8786}}
\affiliation{Experimental Physics Department, CERN, CH-1211 Geneva 23, Switzerland}
\affiliation{KU Leuven, Instituut voor Kern- en Stralingsfysica, B-3001 Leuven, Belgium}

\author{A. A. Breier\orcidlink{0000-0003-1086-9095}}
\affiliation{Institut f\"ur Optik und Atomare Physik, Technische Universit\"at Berlin, 10623 Berlin, Germany}
\affiliation{Laboratory for Astrophysics, Institute of Physics, University of Kassel, Kassel 34132, Germany}

\author{{O. Ahmad}}
\affiliation{KU Leuven, Instituut voor Kern- en Stralingsfysica, B-3001 Leuven, Belgium}

\author{M. Au\orcidlink{0000-0002-8358-7235}}
\affiliation{Systems Department, CERN, CH-1211 Geneva 23, Switzerland}
\affiliation{Department of Chemistry, Johannes Gutenberg-Universit\"{a}t Mainz, 55099 Mainz, Germany}

\author{{S. W.~Bai}}
\affiliation{School of Physics and State Key Laboratory of Nuclear Physics and Technology, Peking University, Beijing 100971, China}

\author{I. Belo\v{s}evi\'{c}}
\affiliation{TRIUMF, Vancouver BC V6T 2A3, Canada}

\author{{J.~Berbalk}}
\affiliation{KU Leuven, Instituut voor Kern- en Stralingsfysica, B-3001 Leuven, Belgium}

\author{R. Berger\orcidlink{0000-0002-9107-2725}}
\affiliation{Fachbereich Chemie, Philipps-Universit\"{a}t Marburg, Marburg 35032, Germany}

\author{{C.~Bernerd}\orcidlink{0000-0002-2183-9695}}
\affiliation{Systems Department, CERN, CH-1211 Geneva 23, Switzerland}

\author{M. L. Bissell}
\affiliation{Department of Physics and Astronomy, The University of Manchester, Manchester M13 9PL, United Kingdom}

\author{A. Borschevsky\orcidlink{0000-0002-6558-1921}}
\affiliation{Van Swinderen Institute of Particle Physics and Gravity, University of Groningen, Groningen 9712 CP, Netherlands}

\author{A. Brinson}
\affiliation{Department of Physics, Massachusetts Institute of Technology, Cambridge, MA 02139, USA}

\author{K. Chrysalidis}
\affiliation{Systems Department, CERN, CH-1211 Geneva 23, Switzerland}

\author{T. E. Cocolios\orcidlink{0000-0002-0456-7878}}
\affiliation{KU Leuven, Instituut voor Kern- en Stralingsfysica, B-3001 Leuven, Belgium}

\author{R. P. de Groote\orcidlink{0000-0003-4942-1220}}
\affiliation{KU Leuven, Instituut voor Kern- en Stralingsfysica, B-3001 Leuven, Belgium}

\author{A. Dorne}
\affiliation{KU Leuven, Instituut voor Kern- en Stralingsfysica, B-3001 Leuven, Belgium}

\author{C. M. Fajardo-Zambrano\orcidlink{0000-0002-6088-6726}}
\affiliation{KU Leuven, Instituut voor Kern- en Stralingsfysica, B-3001 Leuven, Belgium}

\author{R. W. Field\orcidlink{0000-0002-7609-4205}}
\affiliation{Department of Chemistry, Massachusetts Institute of Technology, Cambridge, MA 02139, USA}

\author{K. T. Flanagan\orcidlink{0000-0003-0847-2662}}
\affiliation{Department of Physics and Astronomy, The University of Manchester, Manchester M13 9PL, United Kingdom}
\affiliation{Photon Science Institute, The University of Manchester, Manchester M13 9PY, United Kingdom}

\author{S. Franchoo}
\affiliation{Laboratoire Ir\`{e}ne Joliot-Curie, Orsay F-91405, France}
\affiliation{University Paris-Saclay, Orsay F-91405, France}

\author{R. F. Garcia Ruiz}
\affiliation{Department of Physics, Massachusetts Institute of Technology, Cambridge, MA 02139, USA}
\affiliation{Laboratory for Nuclear Science, Massachusetts Institute of Technology, Cambridge, MA 02139, USA}

\author{K. Gaul}
\affiliation{Fachbereich Chemie, Philipps-Universit\"{a}t Marburg, Marburg 35032, Germany}

\author{S. Geldhof\orcidlink{0000-0002-1335-3505}}
\affiliation{KU Leuven, Instituut voor Kern- en Stralingsfysica, B-3001 Leuven, Belgium}

\author{T. F. Giesen}
\affiliation{Laboratory for Astrophysics, Institute of Physics, University of Kassel, Kassel 34132, Germany}

\author{D. Hanstorp\orcidlink{0000-0001-6490-6897}}
\affiliation{Department of Physics, University of Gothenburg, Gothenburg SE-41296, Sweden}

\author{R. Heinke}
\affiliation{Systems Department, CERN, CH-1211 Geneva 23, Switzerland}

\author{{P. Imgram}\orcidlink{0000-0002-3559-7092}}
\affiliation{KU Leuven, Instituut voor Kern- en Stralingsfysica, B-3001 Leuven, Belgium}

\author{T.~A. Isaev\orcidlink{0000-0003-1123-8316}}
\affiliation{Affiliated with an institute covered by a cooperation agreement with CERN.}

\author{A.~A. Kyuberis\orcidlink{0000-0001-7544-3576}}
\affiliation{Van Swinderen Institute of Particle Physics and Gravity, University of Groningen, Groningen 9712 CP, Netherlands}

\author{S. Kujanp\"{a}\"{a}\orcidlink{0000-0002-5709-3442}}
\affiliation{Department of Physics, University of Jyv\"{a}skyl\"{a}, Jyv\"{a}skyl\"{a} FI-40014, Finland}

\author{L. Lalanne}
\affiliation{KU Leuven, Instituut voor Kern- en Stralingsfysica, B-3001 Leuven, Belgium}
\affiliation{Experimental Physics Department, CERN, CH-1211 Geneva 23, Switzerland}

\author{{P.~Lass\`egues}}
\affiliation{KU Leuven, Instituut voor Kern- en Stralingsfysica, B-3001 Leuven, Belgium}

\author{{J.~Lim}\orcidlink{0000-0002-1803-4642}}
\affiliation{Blackett Laboratory, Centre for Cold Matter, Imperial College London, SW7 2AZ London, United Kingdom}

\author{{Y. C.~Liu}}
\affiliation{School of Physics and State Key Laboratory of Nuclear Physics and Technology, Peking University, Beijing 100971, China}

\author{{K.~M.~Lynch}\orcidlink{0000-0001-8591-2700}}
\affiliation{Department of Physics and Astronomy, The University of Manchester, Manchester M13 9PL, United Kingdom}

\author{{A.~McGlone}\orcidlink{0000-0003-4424-865X}}
\affiliation{Department of Physics and Astronomy, The University of Manchester, Manchester M13 9PL, United Kingdom}

\author{{W. C. Mei}}
\affiliation{School of Physics and State Key Laboratory of Nuclear Physics and Technology, Peking University, Beijing 100971, China}

\author{G. Neyens\orcidlink{0000-0001-8613-1455}}
\email{gerda.neyens@kuleuven.be}
\affiliation{KU Leuven, Instituut voor Kern- en Stralingsfysica, B-3001 Leuven, Belgium}

\author{M. Nichols}
\affiliation{Department of Physics, University of Gothenburg, Gothenburg SE-41296, Sweden}

\author{{L.~Nies}\orcidlink{0000-0003-2448-3775}}
\affiliation{Experimental Physics Department, CERN, CH-1211 Geneva 23, Switzerland}

\author{L. F. Pa\v{s}teka\orcidlink{0000-0002-0617-0524}}
\affiliation{Van Swinderen Institute of Particle Physics and Gravity, University of Groningen, Groningen 9712 CP, Netherlands}
\affiliation{Department of Physical and Theoretical Chemistry, Faculty of Natural Sciences, Comenius University, Bratislava, Slovakia}

\author{H. A. Perrett}
\affiliation{Department of Physics and Astronomy, The University of Manchester, Manchester M13 9PL, United Kingdom}

\author{{A.~Raggio}\orcidlink{0000-0002-5365-1494}}
\affiliation{Department of Physics, University of Jyv\"{a}skyl\"{a}, Jyv\"{a}skyl\"{a} FI-40014, Finland}

\author{J.~R.~Reilly}
\affiliation{Department of Physics and Astronomy, The University of Manchester, Manchester M13 9PL, United Kingdom}

\author{S.~Rothe\orcidlink{0000-0001-5727-7754}}
\affiliation{Systems Department, CERN, CH-1211 Geneva 23, Switzerland}

\author{{E.~Smets}}
\affiliation{KU Leuven, Instituut voor Kern- en Stralingsfysica, B-3001 Leuven, Belgium}

\author{S.-M. Udrescu}
\affiliation{Department of Physics, Massachusetts Institute of Technology, Cambridge, MA 02139, USA}

\author{B. van den Borne\orcidlink{0000-0003-3348-7276}}
\affiliation{KU Leuven, Instituut voor Kern- en Stralingsfysica, B-3001 Leuven, Belgium}

\author{Q. Wang}
\affiliation{School of Nuclear Science and Technology, Lanzhou University, Lanzhou 730000, China}

\author{{J. Warbinek}}
\affiliation{GSI Helmholtzzentrum f\"ur Schwerionenforschung GmbH, 64291 Darmstadt, Germany}
\affiliation{Department of Chemistry - TRIGA Site, Johannes Gutenberg-Universit\"at Mainz, 55128 Mainz, Germany}

\author{J. Wessolek\orcidlink{0000-0001-9804-5538}}
\affiliation{Department of Physics and Astronomy, The University of Manchester, Manchester M13 9PL, United Kingdom}
\affiliation{Systems Department, CERN, CH-1211 Geneva 23, Switzerland}

\author{X. F. Yang}
\affiliation{School of Physics and State Key Laboratory of Nuclear Physics and Technology, Peking University, Beijing 100971, China}

\author{C. Z\"{u}lch}
\affiliation{Fachbereich Chemie, Philipps-Universit\"{a}t Marburg, Marburg 35032, Germany}

\author{the ISOLDE Collaboration}

\date{\today}

\begin{abstract}
We report the spectroscopy of the 14 lowest excited electronic states in the radioactive molecule radium monofluoride (RaF). The observed excitation energies are compared with fully relativistic state-of-the-art Fock-space coupled cluster (FS-RCC) calculations, which achieve an agreement of {$\geq99.64\%$} (within $\sim${12}~meV) with experiment for all states. Guided by theory, a firm assignment of the angular momentum and term symbol is made for 10 states and a tentative assignment for 4 states. The role of high-order electron correlation and quantum electrodynamics effects in the excitation energy of excited states is studied, found to be important for all states. Establishing the simultaneous accuracy and precision of calculations is an important step for research at the intersection of particle, nuclear, and chemical physics, including searches of physics beyond the Standard Model, for which RaF is a promising probe.
\end{abstract}


\maketitle
\section{Introduction}

Achieving high-performance fully relativistic \textit{ab initio} calculations of heavy, many-electron molecules with very high precision and accuracy is a critical milestone for several research areas at the intersection of particle, nuclear, atomic, and molecular physics. This is particularly important for elements that are difficult to study experimentally or for properties that are non-observable in the laboratory. Such research areas include, among others, superheavy element research~\cite{Smits2023SHE}, actinide chemistry~\cite{KovacsGagliardiKonings2015,Rai2024Actinides}, and searches for new physics with heavy molecules~\cite{Safronova2018,Opportunities2024Progressb}, which primarily rely upon the production of radioactive elements at dedicated accelerator facilities.

To assess the performance of \textit{ab initio} computational chemistry, benchmarks using experimental measurements across a wide range of observables are of high importance. Laser spectroscopy offers a powerful avenue to study the electronic structure of radioactive atoms~\cite{Block2021Actinides}, and molecules containing heavy atoms like those in the actinide series~\cite{Heaven2006,Heaven2014}. However, for molecules 
that contain atoms of the seventh period of the periodic table
-- with the exception of the quasi-stable $^{238}$U and $^{232}$Th -- the availability of experimental measurements is significantly hindered by the radioactivity of the heavy nucleus.

As the last pre-actinide element, the electronic structure of radium compounds offers a powerful testing ground for the performance of \textit{ab initio} quantum chemistry. For these molecules, calculations using a fully relativistic Hamiltonian with high-order corrections and the inclusion of electron-correlation effects can be studied with up to a single, non-bonding valence electron.

Radium monofluoride (RaF) in particular has also received a lot of attention due to its promise as a sensitive probe for searches of new physics~\cite{Isaev2010,GarciaRuiz2020}.
Such experiments aim to understand the limitations of the Standard Model (SM) and to assess the validity of candidate theories beyond the SM. To this end, among other approaches, precision tests of the SM and searches for new physics using atomic and molecular spectroscopy are being pursued~\cite{Safronova2018}, for instance to search for the symmetry-violating nuclear Schiff moment~\cite{Flambaum2002,Dobaczewski2018} or the electric dipole moment of the electron (eEDM)~\cite{Roussy2023,Andreev2018,Cesarotti2019b}.

Due to the high degree of precision required by such experiments~\cite{Opportunities2024Progressb}, 
ongoing and future campaigns are focused on systems with maximum sensitivity to the presence of symmetry-violating moments. 
RaF is a particularly promising system as it is amenable to direct laser cooling~\cite{Isaev2010,Udrescu2023}, {which can lead to a further increase in precision by orders of magnitude}~\cite{Fitch2021}. Moreover, the ground state of RaF is highly sensitive to nuclear spin-dependent parity- or time-reversal violation~\cite{Auerbach1996a,Spevak1997,Isaev2010,Petrov2020}, depending on the chosen isotope of the octupole-deformed radium nucleus~\cite{Butler2020}, as well as the eEDM~\cite{Kudashov2014,Sasmal2016,Zhang:2021}.

Extracting values of the symmetry-violating moments from experimental searches requires the calculation of molecular constants that quantify the sensitivity of the molecule to the moment of interest.  Both in atoms and molecules~\cite{Safronova2018,Ginges2004}, the theoretical precision and accuracy of the calculated molecular parameters will dictate the limit to which the symmetry-violating moment can be determined.
As these sensitivity parameters are not experimentally measurable, benchmarking and improving the accuracy and precision of \textit{ab initio} molecular theory across other observables, which can be measured in the laboratory, is also a necessary step towards precision tests of the SM~\cite{Ginges2004,Porsev2010}. Therefore, joint experimental and theoretical efforts have been devoted to evaluating the performance of state-of-the-art \textit{ab initio} methods for many different properties of the structure of RaF.

All isotopes of radium have half-lives from nanoseconds to at most a few days, except for $^{226}$Ra and $^{228}$Ra ({1600 and 5.75 years, respectively}). These two long-lived isotopes have zero nuclear spin and are therefore not suited for the study of symmetry-violating nuclear moments. Radioactive ion beam (RIB) facilities are favorable not only for the preparatory spectroscopic studies needed to understand the electronic structure of the different isotopologues of RaF, but also for future precision experiments. The first spectroscopic studies on RaF molecules were performed at the CERN-ISOLDE radioactive beam facility. This resulted in initial insight into the low-energy electronic-vibrational structure of RaF~\cite{GarciaRuiz2020}, the observation of a strong isotope shift across several short-lived isotopologues~\cite{Udrescu2021}, and a realistic laser-cooling scheme~\cite{Udrescu2023}.

The initial experiment and the interpretation of the data were driven by prior quantum chemistry calculations of the electronic structure of RaF~\cite{GarciaRuiz2020}. Subsequent theoretical studies with single-reference coupled cluster theory including a higher-level treatment of electron correlations and quantum electrodynamic (QED) effects~\cite{Zaitsevskii2022} suggest a re-evaluation of some of the previous spectroscopic assignments~\cite{GarciaRuiz2020}. Furthermore, the very high precision that the calculations achieved for the prediction of {low-lying} excited-state energies, with an uncertainty of only a few tens of cm$^{-1}$ (few meV)~\cite{Zaitsevskii2022,Skripnikov2021}, call for experimental verification of their accuracy. 

This work reports 
the observation of all 14 excited electronic states in RaF that are predicted to exist up to 30,000~cm$^{-1}$ above the ground state. The observed excitation energies are compared to fully relativistic state-of-the-art Fock-space coupled cluster (FS-RCC) calculations~\cite{Eliav2024}, including QED corrections and fully treated triple-cluster amplitudes that capture high-order electron correlation effects. The results highlight the power of FS-RCC for highly precise and accurate calculations of excitation energies in heavy molecules, even at high excitation energies. As a multi-reference approach, the FS-RCC method is also applicable to systems whose states have a multi-reference character~\cite{Zaitsevskii2023ThO,Skripnikov2023AcF}, where single-reference coupled cluster theory is not applicable.

\begin{figure}[h]%
\centering
\includegraphics[width=0.5\textwidth]{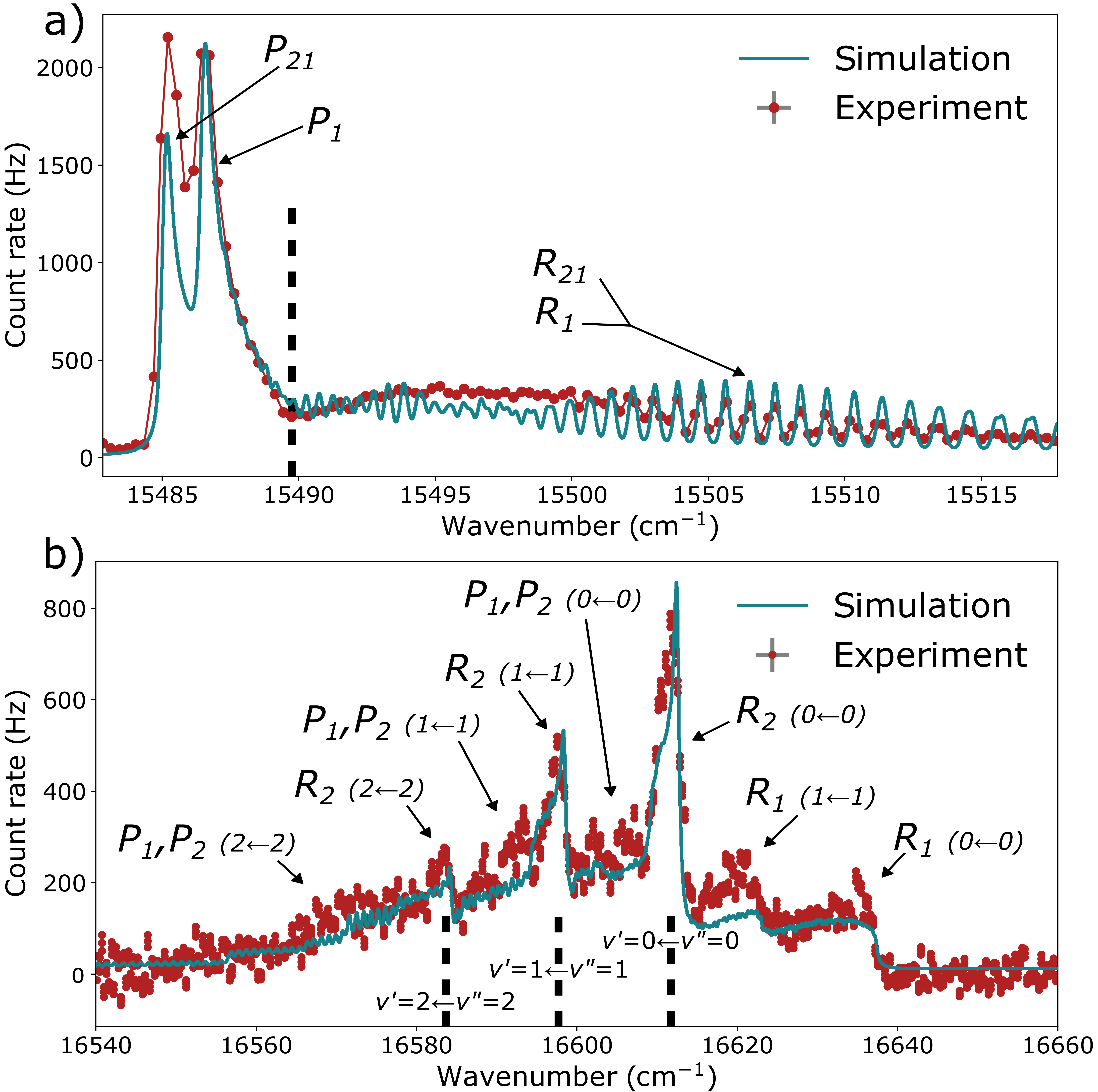}
\caption{Example spectra. \textbf{(a)} $G$~$^2 \Pi_{1/2} \leftarrow A$~$^2 \Pi_{1/2}$ ($v'=0 \leftarrow v''=0$). \textbf{(b)} $C$~$^2 \Sigma_{1/2} \leftarrow X$~$^2 \Sigma_{1/2}$, showing multiple $\Delta v =0$ bands. The simulated spectra were constructed using the best-fit molecular parameters determined from contour fitting with {\sc pgopher}~\cite{Western2017}. The x-axis corresponds to the wavenumber of the scanning laser. {The rotational branches are noted on the plots, and the spectral features marked by a dashed line denote the band heads. Where branch labels lead to the same arrow, the branches cannot be resolved. The parentheses denote the vibrational transition that the rotational branch belongs to.}} \label{fig:setup}
\end{figure}

\section{Results and discussion}
\begin{figure}[h]%
\centering
\includegraphics[width=0.5\textwidth]{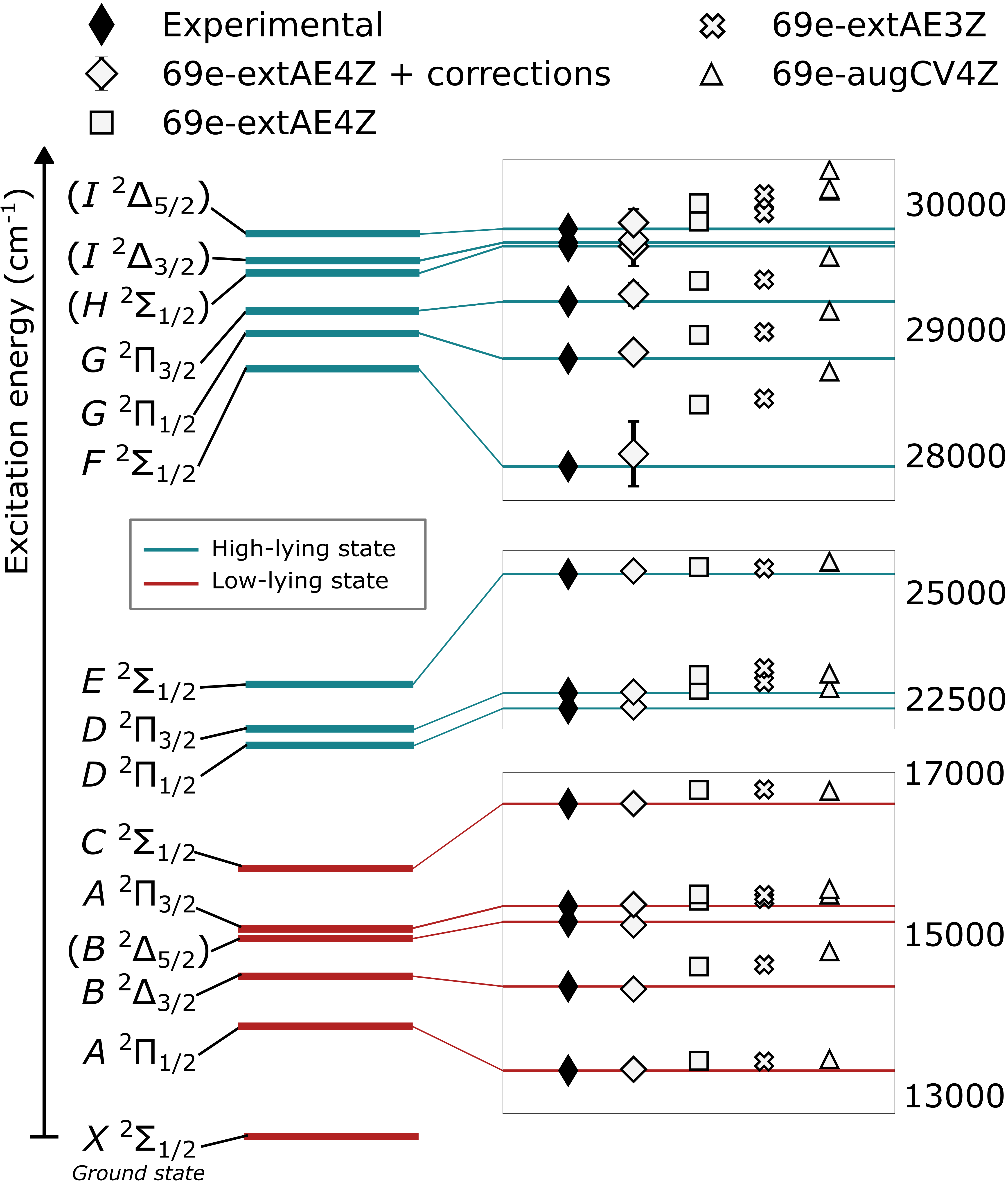}
\caption{\textbf{Left:} Calculated level diagram of RaF up to 30,000~cm$^{-1}$. All levels are observed experimentally {in this work}. The electronic term symbols have been assigned according to the \textit{69e-extAE4Z + corrections} calculations (see text for details). \textbf{Right:} {Zoomed-in} comparison of experimental excitation energies with respect to the vibronic ground state and FS-RCCSD calculations at different levels of sophistication (increasing from right to left). The wavenumber scaling pertains to each inset plot separately. Uncertainties are included only for the most precise calculations (wide diamonds) and in most cases are smaller than the data marker.} \label{fig:diagram}
\end{figure}

Figure~\ref{fig:setup} shows typical experimental spectra obtained in this study, and the rotational branch assignment determined from contour fitting. In Fig.~\ref{fig:diagram}, the experimentally observed excitation energies are compared with the predictions from several FS-RCC calculations at different levels of sophistication.

The search for the initial discovery of the excited states of $^{226}$RaF was guided by theoretical predictions from FS-RCC calculations with single- and double-excitation amplitudes (FS-RCCSD), using doubly augmented (aug) Dyall CV4Z basis sets~\cite{Dyall:09,Dyall:12} and correlating 27 electrons (27e-augCV4Z) within the Dirac-Coulomb Hamiltonian. Such calculations can be completed within a few days
and are therefore well-suited to guide the experimental efforts. Although such calculations have a limited accuracy for high-lying states (within hundreds of cm$^{-1}$) it is sufficient to direct the experimental search to the correct energy range, leading to the first discovery of {10} new excited states (teal lines in Fig.~\ref{fig:diagram} as well as $B$~$^2\Delta_{5/2}$) in this work. 

\begin{figure}[h]%
\centering
\includegraphics[width=0.5\textwidth]{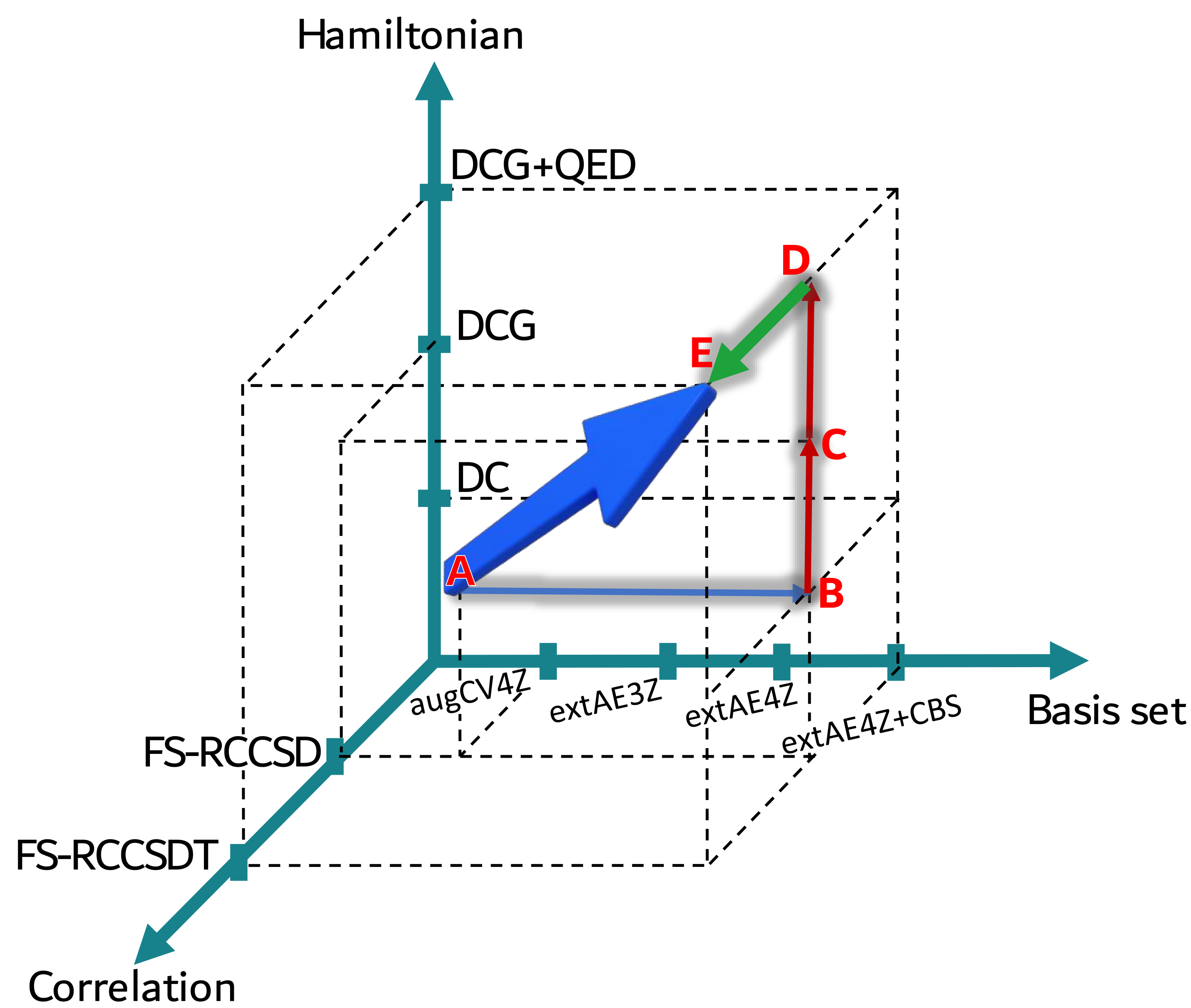}
\caption{Schematic illustration of the electronic calculation scheme. The axes of the coordinate system denote an increase in the basis quality, an improvement in the description of electron correlation effects, and an improvement in the accuracy of the Hamiltonian. Point~E, which represents the most accurate calculations performed in this work, corresponds to the use of the Dirac-Coulomb-Gaunt Hamiltonian incorporating QED corrections (DCG-QED), the use of the extended AE4Z basis set with a basis-set completeness correction (extAE4Z+CBS), and electron correlation effects treated via the fully iterative triple-excitation amplitudes in FS-RCCSDT.} \label{fig:calculation}
\end{figure}

To study the 
accuracy of the electronic structure \textit{ab-initio} methods
as a function of excitation energy, 
further FS-RCCSD calculations were performed at an extended level of correlation treatment, using enhanced basis sets, and an improved electronic Hamiltonian compared to the 27e-augCV4Z calculations that guided the experiment. The improvements in the computational scheme are shown pictorially in Fig.~\ref{fig:calculation}. The agreement between the observed level energies and the most advanced calculations allowed a simplification of the electronic-state assignments.

To improve the treatment of electron correlations, the correlation space was expanded to include 69 electrons (point A in Fig.~\ref{fig:calculation} and triangles in Fig.~\ref{fig:diagram}). Adding the remaining 28 electrons that correspond to the 1$s$--3$d$ shells of Ra, thus including all 97 RaF electrons in the correlation space, modified the level energies by up to 2 cm$^{-1}$ (see Table~III in Supplementary Information), significantly below the total theoretical uncertainty.

To improve the basis-set quality, calculations were performed with the extended (ext) AE3Z~\cite{Skripnikov:2020e,Dyall:2016} (crosses in Fig.~\ref{fig:diagram}) and AE4Z~\cite{Skripnikov2021,Dyall:2016} (squares in Fig.~\ref{fig:diagram}) basis sets, which include a greater number of functions for a more accurate description of the electronic states. A further correction for the incompleteness of the basis sets (CBS correction) was 
implemented based on the scalar-relativistic treatment of valence and outer-core electrons~\cite{Titov:99,Mosyagin:10a,Mosyagin:16} 
(see Supplementary Information for more details), corresponding to point~B in Fig.~\ref{fig:calculation}.

The accuracy of the electronic Hamiltonian  was improved by taking into account the Gaunt inter-electron interaction~\cite{Sikkema:2009} (point C in Fig.~\ref{fig:calculation}) and QED effects~\cite{Shabaev:13} (point D in Fig.~\ref{fig:calculation}), with the latter made possible recently for molecular 4-component calculations~\cite{Skripnikov2021}. Lastly, higher-order electron correlation effects encoded in the triple-excitation amplitudes (T) were included via the FS-RCCSDT approach~\cite{Eliav2024} (point E in Fig.~\ref{fig:calculation}).
The challenging task of simultaneously calculating the triple-excitation contribution to the excitation energies for 15 molecular states that have different electronic configurations was feasible thanks to the use of compact relativistic basis sets~\cite{Skripnikov:13a,Skripnikov2016,Skripnikov:2020e}, developed for use with the 2-component generalized relativistic effective-core potential (GRECP) as the Hamiltonian~\cite{Titov:99,Mosyagin:10a,Mosyagin:16}. The triple-excitation amplitudes were calculated for the 27 outermost electrons (correction denoted as 27e-T), including down to the 5$d$ radium electrons. 

The final theoretical values taking into account CBS, Gaunt, QED, and 27e-T corrections (point~E in Fig.~\ref{fig:calculation}) are included in the wide diamond markers displayed in Fig.~\ref{fig:diagram} and compared to the experimental excitation energies in Table~\ref{tab1}.

An overall agreement of at least {99.64\%} is achieved for all states, which allowed assigning the {newly observed} states and {revising} earlier tentative assignments. A transition observed at 16,175.2(5) cm$^{-1}$ in Ref.~\cite{GarciaRuiz2020} was previously assigned as $C$~$^2\Sigma_{1/2}\leftarrow$ $X$~$^2\Sigma_{1/2}$ $(v'=0\leftarrow v''=0)$. The theoretical precision achieved in the present study together with the new measurements indicate that this transition does not correspond to the lowest electronic excitation energy of the upper state, but rather corresponds to the $(v'=0\leftarrow v''=1)$ vibrational transition, as suggested in Ref.~\cite{Zaitsevskii2022}. Instead, a new spectrum observed in this work very close to the predicted value of 16,615~cm$^{-1}$ is identified as the $v'=0\leftarrow v''=0$ vibronic transition to the $C$~$^2\Sigma_{1/2}$ state.
Additionally, a transition observed at 15,142.7(5)~cm$^{-1}$ in Ref.~\cite{GarciaRuiz2020} was previously tentatively assigned as ($B$~$^2\Delta_{3/2}$)~$\leftarrow X$~$^2\Sigma_{1/2}$ $(v'=0\leftarrow v''=0)$. The close agreement of this transition energy, {observed in this work to lie at 15,140.36(48)[51]~cm$^{-1}$}, with the calculated value of 15,099 cm$^{-1}$ leads to the firm assignment of this transition as ($B$~$^2\Delta_{5/2}$)$\leftarrow X$~$^2\Sigma_{1/2}$ $(v'=0\leftarrow v''=0)$. Finally, a new transition observed at 14,333.00(161)[51]~cm$^{-1}$ is closer to the theoretical prediction for the excitation energy of the $B$~$^2\Delta_{3/2}$ state, and is in agreement with the predictions in Ref.~\cite{Zaitsevskii2022}. Thus, the assignment of the $B$~$^2\Delta_{3/2}$ state at 14,333.00(161)[51]~cm$^{-1}$ is adopted. The term assignments for the {newly observed} high-lying states (teal lines in Fig.~\ref{fig:diagram}), lying above 20,000~cm$^{-1}$ from the ground state, are possible with the highly accurate \textit{ab initio} calculations and elaborated upon in the Supplementary Information.

\begin{figure}[h]%
\centering
\includegraphics[width=0.39\textwidth]{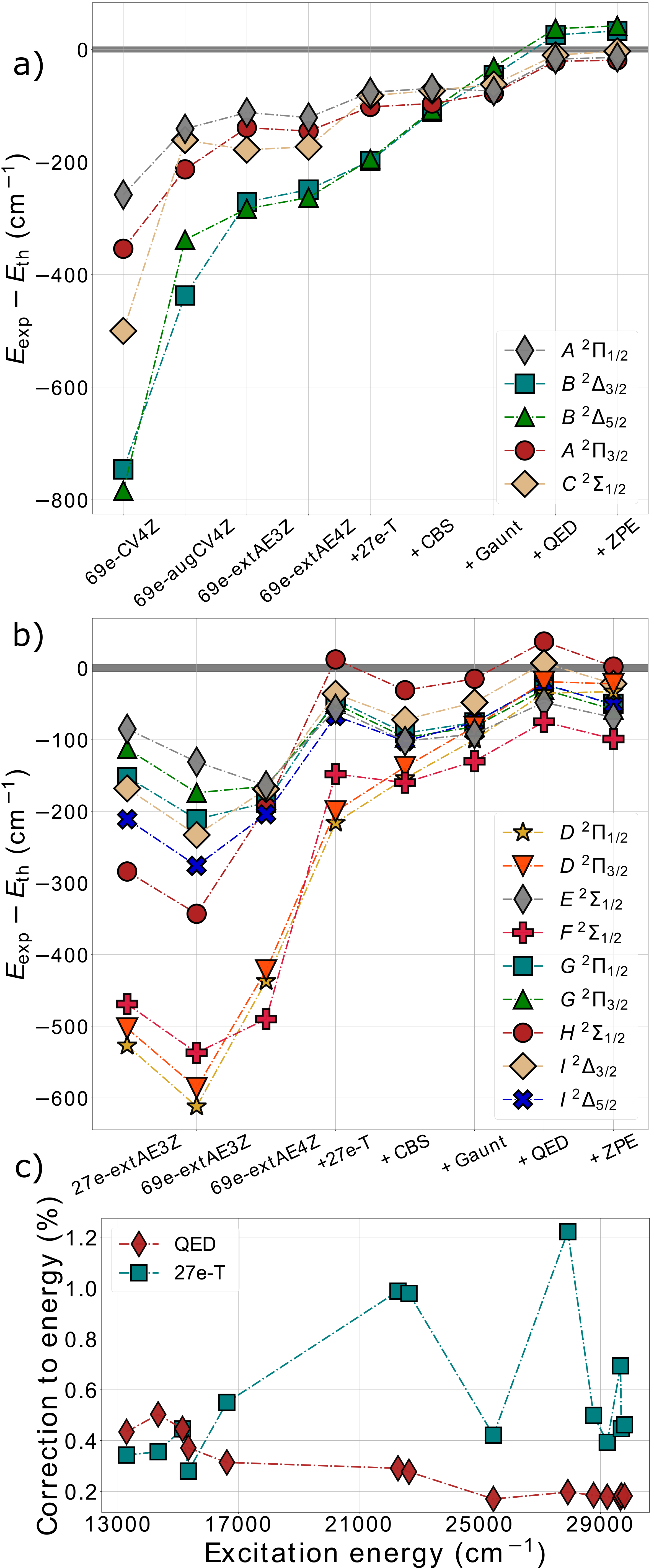}
\caption{Evolution of experiment-theory agreement as a function of increasing theoretical sophistication for (\textbf{a}) the five lowest-lying states and (\textbf{b}) the nine high-lying states. 
For the high-lying states, the CV4Z results did not allow relating calculated and observed levels, and they are not included. 
'+ZPE' corresponds to the zero-point vibrational energy correction. (\textbf{c}) Evolution of the QED and 27e-T corrections to the excitation energies calculated at the 69e-extAE4Z level.} \label{fig:corrections}
\end{figure}
  
Figures~\ref{fig:corrections}a and~\ref{fig:corrections}b present a detailed comparison of the impact of each of the corrections discussed above on the experiment-theory agreement for all states. In particular, the impact of treating triple-excitation amplitudes at high electronic excitation energies is clearly visible in Fig.~\ref{fig:corrections}b. The 27e-T correction has the most prominent effect in improving the agreement with experiment among all listed corrections and for all considered high-lying states. This correction is larger in high-lying states (Fig.~\ref{fig:corrections}b) than in the low-lying ones (Fig.~\ref{fig:corrections}a), demonstrating the need for spectroscopic studies of electronic states far above the ground state to understand the role of correlations in the electronic structure. Figure~\ref{fig:corrections}a also demonstrates the importance of choosing an appropriate basis set for calculations of excited electronic states even energetically close to the ground state, as the difference between 69e-extAE4Z and 69e-CV4Z with the unmodified CV4Z basis sets~\cite{Dyall:09,Dyall:12} is considerable for all states.

\begin{table}[h]
\begin{minipage}{250pt}
\caption{Comparison of experimental and theoretical electronic excitation energies ($T_0$, in cm$^{-1}$) in RaF. The theoretical values correspond to the 69e-extAE4Z calculations with 27e-T, CBS, Gaunt, and QED corrections (wide diamonds in Fig.~\ref{fig:diagram}). The normalized theoretical agreement~($\%$) is reported as $1 - \frac{\lvert E_{\mathrm{th}} - E_{\mathrm{exp}} \rvert}{E_{\mathrm{exp}}}$. Parentheses denote tentative assignment. Statistical and systematic errors are given in round and square brackets {and defined in the Supplementary Information}.}\label{tab1}%
\begin{tabular}{cccc}
\hline \hline
State & Experiment & Theory & Agreement \\ \hline
$X$~$^2 \Sigma_{1/2}$     & 0     & 0                          &      \\
$A$~$^2 \Pi_{1/2}$     & 13,284.427(1)[20]*    & 13,299(36)       & \textbf{99.89}  \\
$B$~$^2 \Delta_{3/2}$ & {14,333.00(161)[51]}   & 14,300(61)  & \textbf{99.77} \\
($B$~$^2 \Delta_{5/2}$) & {15,140.36(48)[51]}   & 15,099(70) & \textbf{99.73}    \\
$A$~$^2 \Pi_{3/2}$    & {15,335.73(49)[62]}  & 15,355(35)  & \textbf{99.87}     \\
$C$~$^2 \Sigma_{1/2}$ & {16,612.06(18)[51]}  & 16,615(69)  & \textbf{99.98} \\
{$D$~$^2 \Pi_{1/2}$} & {22,289.47(29)[51]}  & 22,320(169)& \textbf{99.86}\\
{$D$~$^2 \Pi_{3/2}$} & {22,651.09(31)[51]}  & 22,673(170)  & \textbf{99.90}\\
$E$~$^2 \Sigma_{1/2}$ & 25,451.12(11)[26]  & 25,520(84)  & \textbf{99.73}      \\
{$F$~$^2 \Sigma_{1/2}$} & {27,919.57(180)[51]}  & 28,019(257)&\textbf{99.64}\\
$G$~$^2 \Pi_{1/2}$    & 28,774.07(51)[35]  & 28,824(111) & \textbf{99.83}     \\
$G$~$^2 \Pi_{3/2}$    & 29,225.64(25)[51]  & 29,284(90) & \textbf{99.80}    \\
($H$~$^2 \Sigma_{1/2}$) & 29,665.54(67)[51]  & 29,663(156) & \textbf{99.99}    \\
($I$~$^2 \Delta_{3/2}$) & 29,693.15(24)[51]  & 29,715(102) & \textbf{99.92}\\
($I$~$^2 \Delta_{5/2}$) & 29,801.59(7)[35]  & 29,852(106) & \textbf{99.83}\\
\hline \hline
\end{tabular}
\begin{flushleft}
* Value from Ref.~\cite{Udrescu2023}.\\
\end{flushleft}
\end{minipage}
\end{table}

{In Fig.~\ref{fig:corrections}b, the increase of the correlation space from 27 to 69 electrons using the same basis set leads to an apparent decrease in the agreement with experiment for all states. This can be attributed to a case of mutually cancelling errors when a smaller correlation space is used.}

Finally, in Fig.~\ref{fig:corrections}c the impact of including QED contributions and an iterative treatment of triple-excitation amplitudes is presented.
The contribution of QED effects was found to be especially important for the low-lying states -- excitation energies up to 20,000 cm$^{-1}$ -- having a greater effect on improving the agreement with experiment than the iterative treatment of triple-excitation amplitudes in the FS-RCC models.

Figure~\ref{fig:corrections}c confirms that the relative impact of QED effects is indeed significant at low energies, but it decreases in importance as the average distance of the valence electrons from the radium nucleus increases for greater excitation energies. On the other hand, the higher-order electron correlation effects captured by the iterative treatment of triple-excitation amplitudes are of increasing importance for higher-energy states, but remain non-negligible energetically close to the ground state. This is explained by the participation of non-valence outer-core electrons in the higher-energy excitations. Specifically for the excitation energies of the high-lying states, it is found that the contribution of 5$d$ electrons plays an important role (see Table III in Supplementary Information). {The 27e-T correction is most pronounced particularly for the $D$ and $F$ states,  where it reaches the 1$\%$ level.}

\begin{figure}[h]%
\centering
\includegraphics[width=0.50\textwidth]{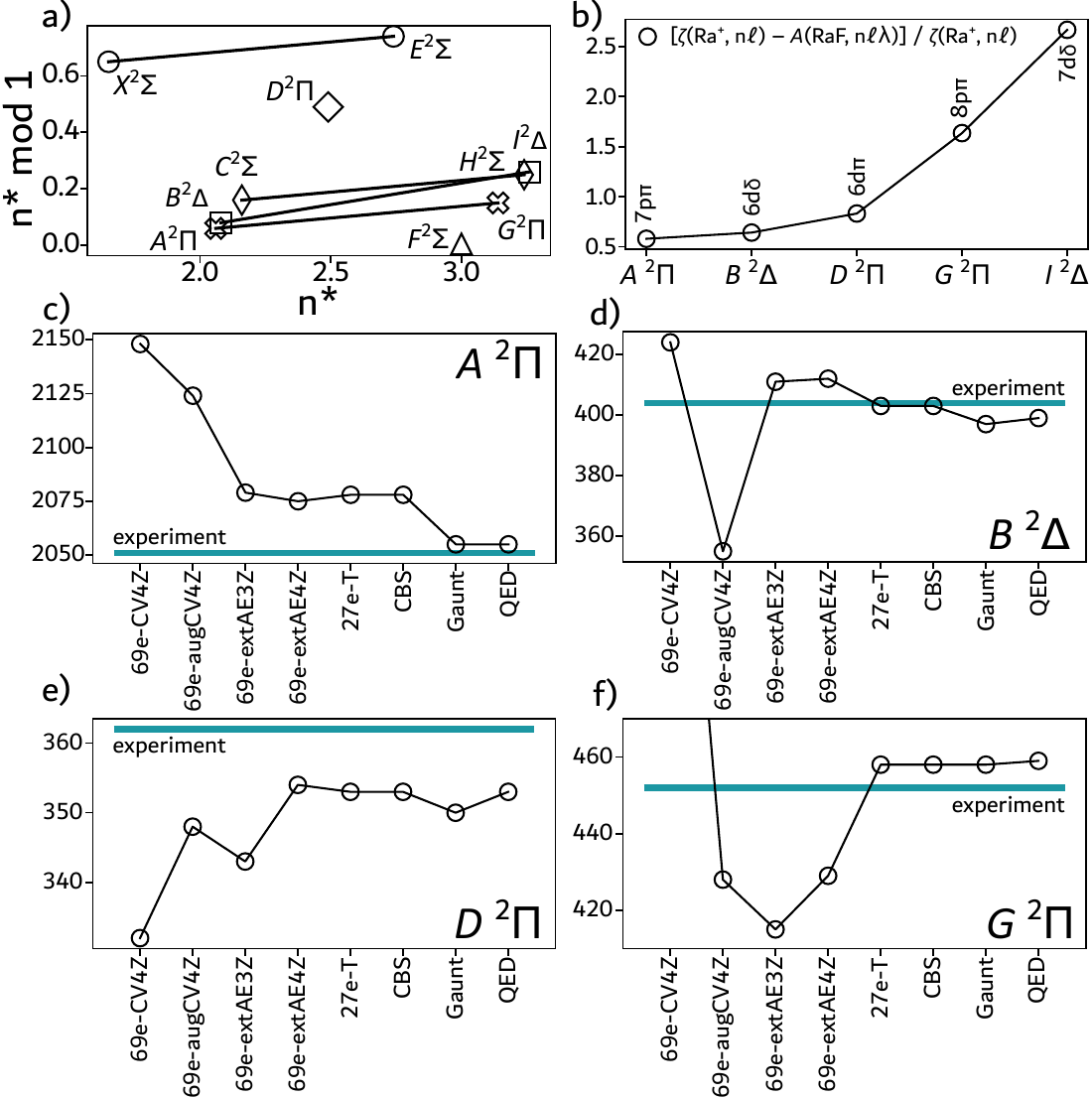}
\caption{
\textbf{(a)} Effective principal quantum numbers $n^*$ and their residuals modulo 1 for excited states in RaF. \textbf{(b)} Normalized difference in SO constant between states in RaF and Ra$^+$ with corresponding (leading) configurations. \textbf{(c)-(f)}
Calculated and observed SO constants (in cm$^{-1}$) for states in RaF as a function of theoretical corrections.} \label{fig:spin_orbit}
\end{figure}

Using the measured ionization potential (IP) of RaF at 4.969(2)$_{\rm{stat}}$(10)$_{\rm{syst}}$~eV~\cite{Wilkins2024RaFIP}, the effective principal quantum number $n^*$ can be extracted for every state in RaF as:
\begin{equation}
    n^* = \sqrt{\frac{R}{{\rm{IP}}-{E}}}
\end{equation}
where $R$ is the Rydberg constant, $E$ is the excitation energy of the state, and $n^*=n-\mu$ with $n$ being the integer principal quantum number and $\mu$ the quantum defect. The $n^*$ values are given in Table~\ref{tab:SO}, and a plot of $n^*\rm{mod}(1)$ versus $n^*$ is shown in Fig.~\ref{fig:spin_orbit}a. Similar to BaF, a total of 10 core-penetrating Rydberg series are expected in RaF -- four $^2\Sigma$, three $^2\Pi$, two $^2\Delta$, and one $^2\Phi$ series -- arising from the mixture of the $s$, $p$, $d$, and $f$ Rydberg series in Ra$^+$ into a core-penetrating, strongly $l$-mixed $s$$\sim$$p$$\sim$$d$$\sim$$f$ supercomplex~\cite{Jakubek2001}. In Fig.~\ref{fig:spin_orbit}a, lines connect states that belong to the same Rydberg series, based on having the same $\Lambda$ and similar $n^*\rm{mod}(1)$~\cite{Jakubek1994Rydberg}. The start of four Rydberg series are identified, two $^2\Sigma$, one $^2\Pi$, and one $^2\Delta$, with two more series starting with $D$~$^2\Pi$ and $F$~$^2\Sigma$. Observing all Rydberg series in the $s$$\sim$$p$$\sim$$d$$\sim$$f$ supercomplex requires further spectroscopy at higher excitation energy in the future.

The bond in alkaline-earth monohalides is well-understood~\cite{Rice1985,Knight1971} to be the result of the electrostatic interaction of a metal cation and a halogen anion, with the valence molecular electron being strongly localized on the metal cation. In RaF, the electric field of F$^-$ polarizes the $^2L$ states of Ra$^+$, causing $\Delta L=\pm1$, $\Delta \lambda=0$ mixing, resulting in $L,\lambda$-dependent energy shifts and Ra$^+$-to-RaF modification of spin-orbit (SO) constants. These effects are dependent on Ra$^+$ electric dipole transition moments, the internuclear distance, and $\Delta L=\pm1$ energy differences of the free Ra$^+$ ion. These polarization effects are responsible for level shifts, level splittings, and configuration mixing, which could be captured by a de-perturbation model for all radium monohalides~\cite{Rice1985}.

\begin{table}[h]
\begin{minipage}{240pt}
\caption{Observed spin-orbit interaction constants $A$ and effective principal quantum numbers $n^*$ for the states in RaF assigned in this work. The constants for $B$~$^2\Delta$ and $I$~$^2\Delta$ are tentative and shown in brackets.}\label{tab:SO}%
\begin{tabular}{lcc}
\hline\hline
         & $A$ (cm$^{-1}$) & $n^*$\\
        \hline
$X$ $^2\Sigma$     &  & 1.65\\
$A$ $^2\Pi$  \hspace{0.5cm}    &  2051(1) & 2.06 \\
$B$ $^2\Delta$  & [404(1)]  & 2.08\\
$C$ $^2\Sigma$     &  & 2.16\\
$D$ $^2\Pi$ & 362(1)  & 2.50\\
$E$ $^2\Sigma$     &  & 2.74\\
$F$ $^2\Sigma$     &  & 3.00\\
$G$ $^2\Pi$  & 452(1) & 3.15\\
$H$ $^2\Sigma$     &  & 3.25\\
$I$ $^2\Delta$ & [54(1)] & 3.26\\
\hline\hline
\end{tabular}
\end{minipage}
\end{table}

The observed SO constants are shown in Table~\ref{tab:SO}. The SO constants for the $A$ and $G$ states as well as the $B$ and $I$ states, which are part of the same $^2\Pi$ and $^2\Delta$ Rydberg series, exhibit reasonable agreement with the expected $(n^{*})^{-3}$ scaling~\cite{Jakubek1994Rydberg}. The SO constants in RaF differ from those in Ra$^+$ due to the field-induced mixing. This effect increases with excitation energy, as seen in Fig.~\ref{fig:spin_orbit}b, because of the increase in the density of Ra$^+$ electronic states. The \textit{ab initio} calculations reflect this energy-dependent configuration mixing, shown as an increase in the number of configurations that contribute $>$10$\%$ to the composition of electronic states above $C$~$^2\Sigma_{1/2}$ (see Table~VI in the Supplementary Information).

In terms of the overall computational precision in this work, it is noted that the absolute difference between theory and experiment in Table~\ref{tab1} is significantly smaller than the theoretical uncertainty for all states, which was determined with an established uncertainty-estimation scheme. The experimental benchmark in Table~\ref{tab1} highlights that the scheme is conservative and the theoretical uncertainty derived from it corresponds more closely to a 2$\sigma$ error. If the agreement between experiment and theory in Table~\ref{tab1} is interpreted as a measure of the theoretical precision -- that is, a quantification of the true theoretical uncertainty -- then the true 1$\sigma$ error of the calculations amounts to $\leq$0.2$\%$ of the excitation energy for all states. This highlights the high precision that FS-RCC calculations can achieve.

A highly accurate and precise treatment of electron correlation in RaF and its reliable uncertainty estimation may also be important for the efforts to calculate the sensitivity of molecular electronic states to nuclear, hadronic, and leptonic symmetry-violating moments, where no experimental verification of computed molecular parameters is possible. All previous theoretical studies of the sensitivity to different symmetry-violating moments in RaF~\cite{Kudashov2014,Sasmal2016,Gaul2017,Gaul2019,Gaul2020toolbox,Zhang:2021} have reported results either using CCSD theory (with triple-excitation amplitudes included only via approximations in some works), or using the Zeroth-Order Regular Approximation based on a mean-field approach and density functional theory, which do not fully capture correlation and relativistic effects. 

Figure~\ref{fig:diagram} and Table~\ref{tab1} show that the 69e-extAE4Z calculations with 27e-T, CBS, Gaunt, and QED corrections reproduce all experimentally observed energies with a deviation better than {0.5$\%$}, which surpasses that of all previous relativistic FS-RCCSD calculations of alkaline-earth monofluorides (Ref.~\cite{Osika2022,Skripnikov:2021b,Hao2019}), while the calculated excitation energies have an uncertainty of $<$1$\%$. Therefore, simultaneously high accuracy and precision is reached even for states energetically far from the ground state.

The achieved agreement justifies the assigned angular momenta and term symbols for the observed levels, leading to a significantly expanded electronic map of RaF. The number of firmly assigned states in this work advances the understanding of the electronic structure of RaF to be on par with that of uranium and thorium molecules~\cite{Gagliardi2005UO2,Antonov2013,Heaven2014}, whose experimental study is not constrained by scarcity or radioactivity. Extensions of the present work to search for states above 30,000~cm$^{-1}$ from the ground state can allow the further assignment of the electronic states in RaF into Rydberg series, which will elucidate the evolution of quantum defects and core-valence electron interactions across the alkaline-earth monofluorides~\cite{Jakubek2001,Kay2008}.

The present study, both experimental and theoretical, paves the way for future high-resolution studies of these states and tests the predictive power of the calculations, whose reliability is a prerequisite for future precision tests of the SM and other areas of fundamental and applied science. Moreover, the performance of FS-RCC in this work highlights the method's suitability for predictions of electronic transitions in wide range of species, such as superheavy atoms and polyatomic molecules, where accurate and precise predictions of the excitation energies are critical for the successful discovery of electronic transitions.

\section{Methods}
\subsection{Experiment}
Laser spectroscopy of $^{226}$RaF was performed using the Collinear Resonance Ionization Spectroscopy (CRIS) experiment at CERN-ISOLDE. 


Accelerated beams of $^{226}$RaF$^+$ were produced at the CERN-ISOLDE radioactive ion beam facility~\cite{Catherall2017}. Two weeks prior to the experiment, short- and long-lived radioactive isotopes, among which $^{226}$Ra nuclei ($t_{1/2}=1600$~years), were produced by impinging 1.4-GeV protons onto a room-temperature uranium carbide target. During the experiment, the irradiated target was gradually heated up to 2000~$^\circ$C to extract the produced radionuclides from within the solid matrix. Due to the asynchronous radionuclide production and extraction from the target, which was chosen so as to suppress beam contaminants with short half-lives, the extraction rate for a given temperature was gradually decreasing over time. At regular intervals, the target temperature was increased in a controlled manner to recover a consistent extraction rate.

By exposing the target to a constant flow of CF$_4$, the radium atoms formed $^{226}$Ra$^{19}$F molecules that were ionized using a rhenium surface ion source. The $^{226}$RaF$^+$ ions were then accelerated to 40~keV and mass-separated from other radiogenic species using two dipolar magnetic separators. The continuous, isotopically pure beam of $^{226}$RaF$^+$ was then accumulated in a radiofrequency quadrupolar cooler-buncher (RFQcb), which released the $^{226}$RaF$^+$ beam at a kinetic energy of 39,900~eV in bunches with a 5-$\mu$s temporal spread once every 20~ms.

\begin{figure}[h]%
\centering
\includegraphics[width=0.49\textwidth]{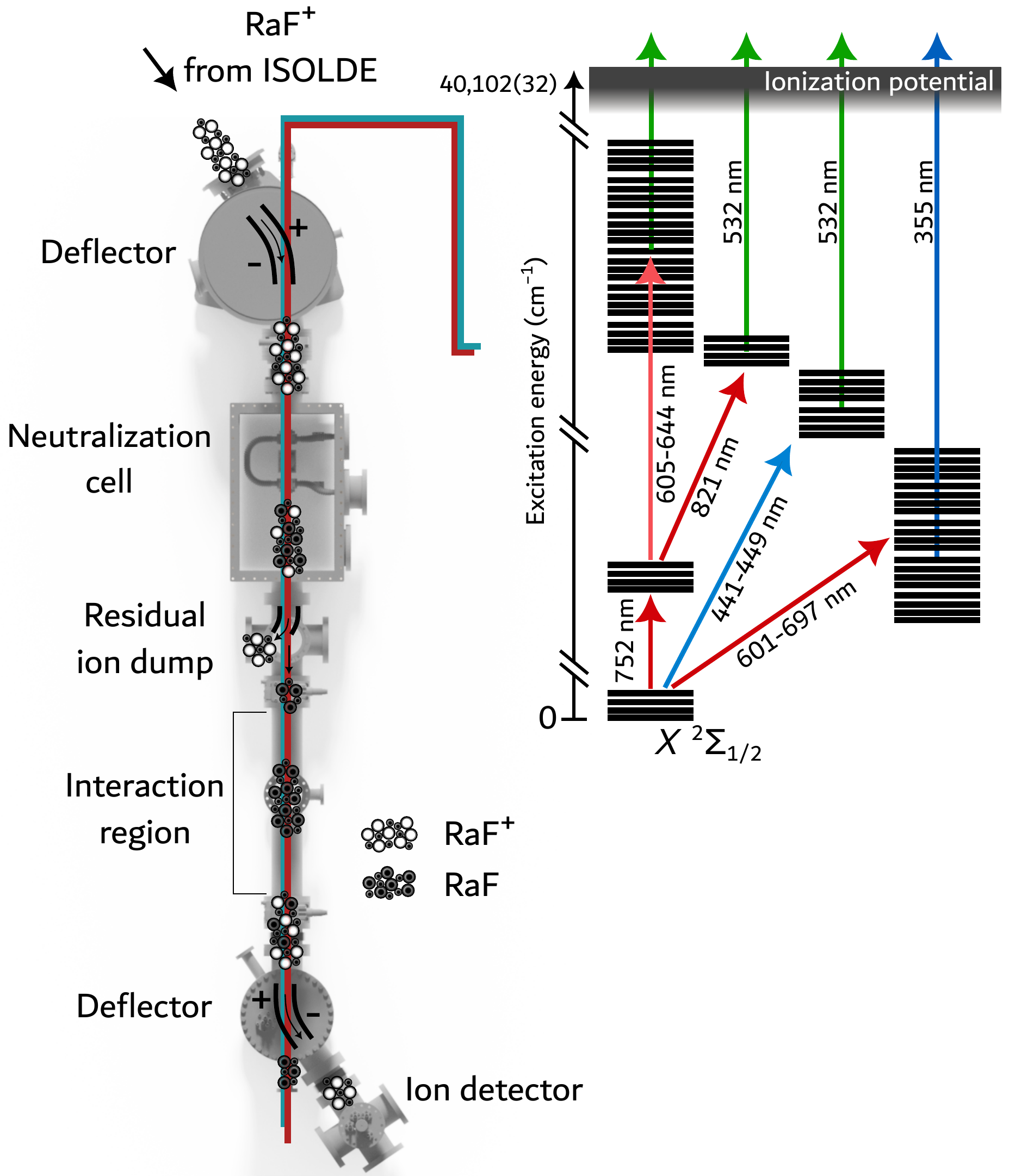}
\caption{Schematic of the CRIS technique and the laser schemes used for resonance ionization spectroscopy of RaF in this work.} \label{fig:CRIS}
\end{figure}

The internal temperature of the beam was cooled to near room temperature while being trapped in the RFQcb in the presence of a helium buffer gas. The temperature at which the ensemble delivered from the RFQcb thermalized was found to vary over time, ranging from 350 to 600~K. In all cases, the spectra could be fitted assuming that the molecular ensemble had a uniform temperature distribution. Therefore, it is considered that above a certain target temperature, the produced molecules could not be cooled to room temperature within the time they were trapped in the RFQcb, but thermalized at a temperature between the target environment and the room-temperature buffer gas.

A typical $^{226}$RaF$^+$ beam intensity of 1.2$\times$10$^{6}$ ions per second was then sent into the CRIS beam line, which is shown schematically in Fig.~\ref{fig:CRIS}. At CRIS, the fast ion beam was firstly neutralized via collisions with hot sodium vapor in a charge-exchange cell, at the exit of which all the ions that were not successfully neutralized were deflected onto a beam dump. The bunches of neutralized molecules entered the interaction region, where they interacted collinearly with pulsed laser beams that step-wise excited the molecular electrons from the ground state to above the IP. The final laser step in the CRIS scheme induced non-resonant ionization of the molecules that were resonantly brought to an excited state by the preceding laser steps. As a result, successful ionization of the neutralized molecules required one (in two-step schemes) or two (in three-step schemes) resonant laser excitations. At the end of the interaction region, the resonantly ionized molecules were deflected away from the residual neutral particles and onto a single ion counter using electrostatic deflectors. The molecular spectra were obtained by monitoring the ion count rate as a function of the laser frequencies of the resonant excitation steps.

\subsection{Laser setup} \label{appsubsec:laser_setup}
All lasers used in this work were pulsed and overlapped with the molecular beam in a collinear geometry. Multiple laser schemes were used for the spectroscopy of the excited states in RaF. For each scheme, the molecules underwent either one or two resonant excitations starting from the electronic ground state using tunable  pulsed titanium-sapphire (Ti:Sa) or pulsed dye lasers (PDLs, Spectra Physics PDL and Sirah Cobra). A high-power 532-nm or 355-nm Nd:YAG laser was used to ionize the molecules that had been resonantly excited. The molecular excitation energies were measured by scanning the frequency of the tunable laser used for a resonant transition while monitoring the ion count rate.

The level search was facilitated by preparing multiple laser-ionization schemes based on two PDLs with nominal linewidths~$\Delta f$ of 0.8 and 9~GHz and two grating-based broadband pulsed Ti:Sa lasers ($\Delta f \approx 3$~GHz). A total wavenumber range $>$4,000~cm$^{-1}$ was scanned in a period of a few days. Further information about the data analysis and spectroscopic assignment can be found in the Supplementary Information. More details on the experimental technique can be found in Refs.~\cite{GarciaRuiz2020,Cocolios2013CRIS}.

The spectrum of the previously reported $A$~$^{2}\Pi_{1/2}$ state was measured 
using a grating-based titanium:sapphire (Ti:Sa) laser scanned around 752~nm (in the molecular rest frame), corresponding to the transition from the ground state. 
The transition from the electronic ground state to this level was then used as the first step in three-step schemes to search for {\color{black}most} high-lying levels, using either its $v'=0 \leftarrow v''=0$ or $v'=1 \leftarrow v''=1$ vibrational transitions.

The spectra of the transitions to the $B$~$^{2}\Delta_{3/2}$, $B$~$^{2}\Delta_{5/2}$, $A$~$^{2}\Pi_{3/2}$, and $C$~$^{2}\Sigma_{1/2}$ states were measured with two-step schemes by scanning a dye laser around 697~nm (Pyridine~1) for $B$~$^{2}\Delta_{3/2}$, 661~nm (DCM) for $B$~$^{2}\Delta_{5/2}$, 651~nm (DCM) for $A$~$^{2}\Pi_{3/2}$, and 601~nm (Pyridine~1) for $C$~$^{2}\Sigma_{1/2}$, followed by a 355-nm non-resonant ionization step. 
The $D$~$^2\Pi$ states were also measured with a two-step scheme using second-harmonic Ti:Sa for the resonant excitation from the $X$~$^2\Sigma_{1/2}$ ground state, followed by non-resonant ionization using a 532-nm Nd:YAG laser.

The remaining high-lying states were observed using three-step schemes, where the first step was the resonant transition from $X$~$^2\Sigma_{1/2}$ to $A$~$^2\Pi_{1/2}$ ($0\leftarrow0$ or $1\leftarrow1$), followed by the resonant excitation to the high-lying states. The ionization was induced by a non-resonant excitation driven by a 532-nm Nd:YAG laser. To search for the transition to the $E$~$^2 \Sigma_{1/2}$ state, the second laser step was scanned around 821~nm with a second grating-based Ti:Sa laser. For the transitions to the $G$~$^{2}\Pi_{1/2}$, $G$~$^{2}\Pi_{3/2}$, $H$~$^{2}\Sigma_{1/2}$, $I$~$^{2}\Delta_{3/2}$, and $I$~$^{2}\Delta_{5/2}$ states, the second laser-excitation step was scanned around 644~nm, 625~nm, 610~nm, 609~nm, and 605~nm, respectively, using a dye laser (DCM). For the transition to the $F$~$^2\Sigma_{1/2}$ state, a dye laser at 683~nm (Pyridine 1) was used. Figure~\ref{fig:CRIS} pictorially summarizes the laser schemes used in this work.

The Ti:Sa lasers were pumped by a 532-nm Nd:YAG laser operating at 1~kHz, while the dye lasers were pumped by 532-nm Nd:YAG lasers operating at 50~Hz. The 532-nm Nd:YAG laser used for non-resonant ionization was operating at 50~Hz, as well, {\color{black}while the 355-nm Nd:YAG laser was operating at 100~Hz}. The relative timing between the laser steps was controlled by triggering the Q-switches of the pulsed lasers using a multi-channel, ultra-low-jitter clock (Quantum Composer 9528). As the excited-state lifetimes were not known for the newly discovered electronic states, all three steps were temporally overlapped. The frequencies of the dye lasers were measured using a HighFinesse WS6-600 wavemeter and the frequencies of the Ti:Sa lasers were measured using a HighFinesse WSU-2 wavemeter. The WSU-2 wavemeter was continuously calibrated by monitoring at the same time the frequencies of a diode laser (Toptica DL pro) locked to a hyperfine peak in rubidium.

\subsection{Calculations} \label{appsubsec:fs-rcc}
FS-RCCSD calculations were performed using the {\sc dirac}~\cite{DIRAC:19,DIRAC:20} code. FS-RCCSDT calculations were performed using the {\sc exp-t} code ~\cite{Oleynichenko2020,EXPT:20}. Single reference relativistic coupled cluster calculations were performed with the~{\sc mrcc}~\cite{Kallay:6,MRCC2020} code. All scalar relativistic correlation calculations were performed using the public version of the {\sc cfour}~\cite{CFOUR} code. Matrix elements of the model QED Hamiltonian were calculated within the code developed in Ref.~\cite{Skripnikov2021}.


The electronic ground state of the RaF$^+$ cation 
was chosen as a Fermi-vacuum in the Fock-space (FS) calculation. Target states in the neutral RaF are considered as belonging to the one-particle sector of the FS. In the calculation, the Dirac–Coulomb Hamiltonian was used to solve the self-consistent (Dirac-Hartree-Fock) problem and then converted to the two-component all-electron Hamiltonian by means of the X2C technique within the molecular mean-field approximation~\cite{Sikkema:2009}.
For the 69-electron FS-RCCSD calculations, the extAE4Z basis set was used, which corresponds to the extended uncontracted Dyall's all-electron AE4Z basis set for Ra~\cite{Dyall:12} from Ref.~\cite{Skripnikov2021} and the uncontracted Dyall's AAEQZ basis set~\cite{Dyall:2016} for F. Explicitly, this basis set includes  $[42s\, 38p\, 27d\, 27f\, 13g\, 3h\, 2i]$ Gaussian-type functions  for Ra. In these FS-RCCSD calculations, the energy cutoff for virtual orbitals was set to 300 Hartree to ensure the proper correlation of outer-core electrons. For the 69-electron FS-RCCSD calculations (employed in the initial discovery of the excited states) using a doubly augmented Dyall's CV4Z basis set~\cite{Dyall:09,Dyall:12}, the energy cutoff for virtual orbitals was set to 100 Hartree.

In addition to the 69e-FS-RCCSD-extAE4Z calculations, further calculations with different numbers of correlated electrons and utilizing different basis sets were performed
to probe various computational aspects. These included 27- and 69-electron calculations in the standard (unmodified) uncontracted Dyall's CV4Z basis set and 17-, 27-, 35-, 69-, and 97-electron calculations in the extended AE3Z basis set~\cite{Dyall:09,Dyall:12,Dyall:2016} (see Tables~III and IV in Supplementary Information). The latter extAE3Z basis set has been developed in Ref.~\cite{Skripnikov:2020e} and includes $[38s\, 33p\, 24d\, 14f\, 7g\, 3h\, 2i]$ Gaussian-type functions for Ra and corresponds to the uncontracted AE3Z~\cite{Dyall:2016} basis set on F.  

To take into account contributions of an extended number of basis functions with high angular momentum ($L\geq 4$) beyond those contained in extAE4Z, the scalar-relativistic variant of the valence part of the generalized relativistic effective core potential approach~\cite{Titov:99,Mosyagin:10a,Mosyagin:16} was used, as well as the 37e-EOMEA-CCSD approach (which is equivalent to FS-RCCSD in the considered case) to treat electron correlations using the {\sc cfour} code~\cite{CFOUR}. In this way, it was possible to extend the basis set towards higher harmonics in Ra up to $15g\, 15h\, 15i$ (with an additional increase in the number of $s\, p\, d$ functions), which is intractable in practice within the Dirac-Coulomb Hamiltonian using available resources. Following Ref.~\cite{Skripnikov2021}, the extrapolated contribution of higher harmonics to the basis-set correction was also included. In the extrapolation scheme, the contribution of basis functions with an angular momentum $L$ (for $L>6$) is determined using the formula $A_1/L^5 + A_2/L^6$, where the coefficients {$A_1$ and $A_2$} were derived from the directly calculated contributions of $h$- ($L$=5) and $i$- ($L$=6) harmonics. Thus, we calculated the sum of contributions for $L>6$. This scheme has been optimized and tested in Ref.~\cite{Skripnikov2021} for Ra$^+$ and RaF excitation energies, and has also been successfully applied to calculated excitation energies in Ba$^+$ and BaF~\cite{Skripnikov:2021b}. The contribution of the increased basis set described above is termed ``+CBS'' in the main text.

Correlation effects beyond the FS-RCCSD model have been calculated as the difference in transition energies calculated within the relativistic FS-RCCSDT and FS-RCCSD approaches using specially constructed compact natural contracted basis sets~\cite{Skripnikov:13a,Skripnikov2016,Skripnikov:2020e}, correlating 27 RaF electrons and employing two-component (valence) GRECP Hamiltonian.

The atomic natural-like compact basis was constructed in such a way as to describe the $7S$, $6D$, $7P$, $8S$, $7D$, $5F$ and $8P$ states of the Ra$^+$ cation, which are relevant for the considered electronic states of RaF. 
For these states in Ra$^+$, scalar-relativistic 37e-CCSD(T) calculations were performed in an extended basis set, yielding correlated one-particle density matrices. 
For a proper treatment of spin-orbit effects, we also calculated the one-particle density matrices corresponding to the occupied and virtual atomic spinors obtained in the two-component Hartree-Fock calculation. In the scalar atomic orbital (AO) basis function representation, these density matrices include not only diagonal ``spin-up'' and ``spin-down'' blocks, as in scalar-relativistic calculations, but also mixed-spin blocks. Density matrices in the AO representation, obtained from both correlation calculations and two-component Hartree-Fock calculations, were averaged over electronic states, spin blocks, and projections of the AO basis functions. The resulting effective density matrix was then used as the matrix of the density operator in the AO representation. Corresponding eigenvectors with the highest eigenvalues (occupation numbers) were employed for constructing the compact basis set used in the RaF calculations.
In principle, such a procedure could also be used within the Dirac-Coulomb Hamiltonian. However, at present one cannot use contracted Gaussian basis sets for heavy elements such as Ra in available implementations of the Dirac-Coulomb Hamiltonian.
Another obstacle to the direct use of the Dirac-Coulomb Hamiltonian for the compact basis-set construction is the presence of  serious practical limitations in the size of the original basis set that is used to construct correlated density matrices. 

Lastly, the contributions of QED~\cite{Shabaev:13,Skripnikov2021} as well as Gaunt inter-electron effects~\cite{Sikkema:2009} were calculated at the FS-RCCSD level.

To obtain the composition of the RaF molecular electronic states in terms of Ra$^+$ states, the expectation values of the projectors on the corresponding Ra$^+$ states were calculated, e.g., $|7s_{1/2}\rangle\langle7s_{1/2}|$, $|7p_{1/2}\rangle\langle7p_{1/2}|$, etc., where the one-electron functions $|7s_{1/2}\rangle$ and $|7p_{1/2}\rangle$ were obtained in the atomic relativistic Hartree-Fock calculation and correspond to the leading configurations in the Ra$^+$ many-electron functions. These composition calculations were performed within the compact basis set described above. For RaF, the FS-RCCSD wavefunctions were used. Due to the intrinsic semi-quantitative character of the procedure used, each contribution was rounded to tens of a percent.

\section*{Data availability}
The binned spectra that were analyzed will be available in the form of {\sc pgopher} overlay files after publication at the reserved doi: \href{https://doi.org/10.5281/zenodo.8196151}{10.5281/zenodo.8196151}. The raw data and analysis code can be provided upon request to the authors following publication. Further information on the experimental methods are provided in the Supplementary Information.

\section*{Author contribution}
M.A.-K., S.G.W., L.V.S., and G.N. led the manuscript preparation. M.A.-K., S.G.W., \'A.K., A.A.B., O.A., M.Au, S.W.B., I.B., J.B., R.B., C.B., M.L.B., A.B., A.Br., K.C., T.E.C., R.P.d.G., A.D., C.M.F.Z., K.T.F., R.W.F., S.F., R.F.G.R., K.G., S.G., T.F.G., D.H., R.H., P.I., S.K., L.L., P.L., J.L., Y.C.L., K.M.L., A.McG., W.C.M., G.N., M.N., L.N., H.A.P., A.R., J.R.R., S.R., E.S., S.-M.U., B.v.d.B., Q.W., J.Wa., J.We., and X.F.Y. performed the experiment. L.V.S. R.B., A.B., K.G., T.A.I., A.A.K., L.F.P., and C.Z. performed the calculations. M.A.-K., A.A.B., and C.M.F.Z. performed the data analysis. All coauthors participated in editing and revising the manuscript.

\section*{Acknowledgments}
We thank the ISOLDE collaboration {for funding support} and the ISOLDE technical teams for their work and assistance.

{This project has received funding from the European Union's Horizon Europe Research and Innovation programme under Grant Agreement No. 101057511. The research leading to these results has received funding from the European Union's Horizon 2020 research and innovation programme under Grant Agreement No. 654002.}

Financial support from FWO, as well as from the Excellence of Science (EOS) programme (No. 40007501) and the KU Leuven project C14/22/104, is acknowledged. The STFC consolidated grants ST/V001116/1 and ST/P004423/1 and the FNPMLS ERC grant agreement 648381 are acknowledged. SGW, RFGR, and SMU acknowledge funding by the Office of Nuclear Physics, U.S. Department of Energy Grants DE-SC0021176 and DE-SC002117. AAB, TFG, RB, and KG acknowledge funding from the Deutsche Forschungsgemeinschaft (DFG) – Projektnummer 328961117 – SFB 1319 ELCH. We thank the Center for Information Technology at the University of Groningen for their support and for providing access to the Peregrine high performance computing cluster. MAu, MN, AR, JWa, and JWe acknowledge funding from the EU’s H2020-MSCA-ITN Grant No. 861198 ‘LISA’. DH acknowledges financial support from the Swedish Research Council (2020-03505). JL acknowledges financial support from STFC grant ST/V00428X/1. LFP acknowledges the support from the Dutch Research Council (NWO) project number VI.C.212.016 and the Slovak Research and Development Agency projects APVV-20-0098 and APVV-20-0127.

\section*{Declarations}
The authors declare no conflict of interest.

%

\end{document}


\title{Supplementary Information: Pinning down electron correlations in RaF via spectroscopy of excited states and high-accuracy relativistic quantum chemistry}

\author{M. Athanasakis-Kaklamanakis\orcidlink{0000-0003-0336-5980}}
 \email{m.athkak@cern.ch}
\affiliation{Experimental Physics Department, CERN, CH-1211 Geneva 23, Switzerland}
\affiliation{KU Leuven, Instituut voor Kern- en Stralingsfysica, B-3001 Leuven, Belgium}
\affiliation{Blackett Laboratory, Centre for Cold Matter, Imperial College London, SW7 2AZ London, United Kingdom}

\author{S. G. Wilkins}
 \email{wilkinss@mit.edu}
\affiliation{Department of Physics, Massachusetts Institute of Technology, Cambridge, MA 02139, USA}
\affiliation{Laboratory for Nuclear Science, Massachusetts Institute of Technology, Cambridge, MA 02139, USA}

\author{L. V. Skripnikov\orcidlink{0000-0002-2062-684X}}
 \email{skripnikov\_lv@pnpi.nrcki.ru}
\affiliation{Affiliated with an institute covered by a cooperation agreement with CERN.}

\author{\'{A}. Koszor\'{u}s\orcidlink{0000-0001-7959-8786}}
\affiliation{Experimental Physics Department, CERN, CH-1211 Geneva 23, Switzerland}
\affiliation{KU Leuven, Instituut voor Kern- en Stralingsfysica, B-3001 Leuven, Belgium}

\author{A. A. Breier\orcidlink{0000-0003-1086-9095}}
\affiliation{Institut f\"ur Optik und Atomare Physik, Technische Universit\"at Berlin, 10623 Berlin, Germany}
\affiliation{Laboratory for Astrophysics, Institute of Physics, University of Kassel, Kassel 34132, Germany}

\author{{O. Ahmad}}
\affiliation{KU Leuven, Instituut voor Kern- en Stralingsfysica, B-3001 Leuven, Belgium}

\author{M. Au\orcidlink{0000-0002-8358-7235}}
\affiliation{Systems Department, CERN, CH-1211 Geneva 23, Switzerland}
\affiliation{Department of Chemistry, Johannes Gutenberg-Universit\"{a}t Mainz, 55099 Mainz, Germany}

\author{{S. W.~Bai}}
\affiliation{School of Physics and State Key Laboratory of Nuclear Physics and Technology, Peking University, Beijing 100971, China}

\author{I. Belo\v{s}evi\'{c}}
\affiliation{TRIUMF, Vancouver BC V6T 2A3, Canada}

\author{{J.~Berbalk}}
\affiliation{KU Leuven, Instituut voor Kern- en Stralingsfysica, B-3001 Leuven, Belgium}

\author{R. Berger\orcidlink{0000-0002-9107-2725}}
\affiliation{Fachbereich Chemie, Philipps-Universit\"{a}t Marburg, Marburg 35032, Germany}

\author{{C.~Bernerd}\orcidlink{0000-0002-2183-9695}}
\affiliation{Systems Department, CERN, CH-1211 Geneva 23, Switzerland}

\author{M. L. Bissell}
\affiliation{Department of Physics and Astronomy, The University of Manchester, Manchester M13 9PL, United Kingdom}

\author{A. Borschevsky\orcidlink{0000-0002-6558-1921}}
\affiliation{Van Swinderen Institute of Particle Physics and Gravity, University of Groningen, Groningen 9712 CP, Netherlands}

\author{A. Brinson}
\affiliation{Department of Physics, Massachusetts Institute of Technology, Cambridge, MA 02139, USA}

\author{K. Chrysalidis}
\affiliation{Systems Department, CERN, CH-1211 Geneva 23, Switzerland}

\author{T. E. Cocolios\orcidlink{0000-0002-0456-7878}}
\affiliation{KU Leuven, Instituut voor Kern- en Stralingsfysica, B-3001 Leuven, Belgium}

\author{R. P. de Groote\orcidlink{0000-0003-4942-1220}}
\affiliation{KU Leuven, Instituut voor Kern- en Stralingsfysica, B-3001 Leuven, Belgium}

\author{A. Dorne}
\affiliation{KU Leuven, Instituut voor Kern- en Stralingsfysica, B-3001 Leuven, Belgium}

\author{C. M. Fajardo-Zambrano\orcidlink{0000-0002-6088-6726}}
\affiliation{KU Leuven, Instituut voor Kern- en Stralingsfysica, B-3001 Leuven, Belgium}

\author{R. W. Field\orcidlink{0000-0002-7609-4205}}
\affiliation{Department of Chemistry, Massachusetts Institute of Technology, Cambridge, MA 02139, USA}

\author{K. T. Flanagan\orcidlink{0000-0003-0847-2662}}
\affiliation{Department of Physics and Astronomy, The University of Manchester, Manchester M13 9PL, United Kingdom}
\affiliation{Photon Science Institute, The University of Manchester, Manchester M13 9PY, United Kingdom}

\author{S. Franchoo}
\affiliation{Laboratoire Ir\`{e}ne Joliot-Curie, Orsay F-91405, France}
\affiliation{University Paris-Saclay, Orsay F-91405, France}

\author{R. F. Garcia Ruiz}
\affiliation{Department of Physics, Massachusetts Institute of Technology, Cambridge, MA 02139, USA}
\affiliation{Laboratory for Nuclear Science, Massachusetts Institute of Technology, Cambridge, MA 02139, USA}

\author{K. Gaul}
\affiliation{Fachbereich Chemie, Philipps-Universit\"{a}t Marburg, Marburg 35032, Germany}

\author{S. Geldhof\orcidlink{0000-0002-1335-3505}}
\affiliation{KU Leuven, Instituut voor Kern- en Stralingsfysica, B-3001 Leuven, Belgium}

\author{T. F. Giesen}
\affiliation{Laboratory for Astrophysics, Institute of Physics, University of Kassel, Kassel 34132, Germany}

\author{D. Hanstorp\orcidlink{0000-0001-6490-6897}}
\affiliation{Department of Physics, University of Gothenburg, Gothenburg SE-41296, Sweden}

\author{R. Heinke}
\affiliation{Systems Department, CERN, CH-1211 Geneva 23, Switzerland}

\author{{P. Imgram}\orcidlink{0000-0002-3559-7092}}
\affiliation{KU Leuven, Instituut voor Kern- en Stralingsfysica, B-3001 Leuven, Belgium}

\author{T.~A. Isaev\orcidlink{0000-0003-1123-8316}}
\affiliation{Affiliated with an institute covered by a cooperation agreement with CERN.}

\author{A.~A. Kyuberis\orcidlink{0000-0001-7544-3576}}
\affiliation{Van Swinderen Institute of Particle Physics and Gravity, University of Groningen, Groningen 9712 CP, Netherlands}

\author{S. Kujanp\"{a}\"{a}\orcidlink{0000-0002-5709-3442}}
\affiliation{Department of Physics, University of Jyv\"{a}skyl\"{a}, Jyv\"{a}skyl\"{a} FI-40014, Finland}

\author{L. Lalanne}
\affiliation{KU Leuven, Instituut voor Kern- en Stralingsfysica, B-3001 Leuven, Belgium}
\affiliation{Experimental Physics Department, CERN, CH-1211 Geneva 23, Switzerland}

\author{{P.~Lass\`egues}}
\affiliation{KU Leuven, Instituut voor Kern- en Stralingsfysica, B-3001 Leuven, Belgium}

\author{{J.~Lim}\orcidlink{0000-0002-1803-4642}}
\affiliation{Blackett Laboratory, Centre for Cold Matter, Imperial College London, SW7 2AZ London, United Kingdom}

\author{{Y. C.~Liu}}
\affiliation{School of Physics and State Key Laboratory of Nuclear Physics and Technology, Peking University, Beijing 100971, China}

\author{{K.~M.~Lynch}\orcidlink{0000-0001-8591-2700}}
\affiliation{Department of Physics and Astronomy, The University of Manchester, Manchester M13 9PL, United Kingdom}

\author{{A.~McGlone}\orcidlink{0000-0003-4424-865X}}
\affiliation{Department of Physics and Astronomy, The University of Manchester, Manchester M13 9PL, United Kingdom}

\author{{W. C. Mei}}
\affiliation{School of Physics and State Key Laboratory of Nuclear Physics and Technology, Peking University, Beijing 100971, China}

\author{G. Neyens\orcidlink{0000-0001-8613-1455}}
\email{gerda.neyens@kuleuven.be}
\affiliation{KU Leuven, Instituut voor Kern- en Stralingsfysica, B-3001 Leuven, Belgium}

\author{M. Nichols}
\affiliation{Department of Physics, University of Gothenburg, Gothenburg SE-41296, Sweden}

\author{{L.~Nies}\orcidlink{0000-0003-2448-3775}}
\affiliation{Experimental Physics Department, CERN, CH-1211 Geneva 23, Switzerland}

\author{L. F. Pa\v{s}teka\orcidlink{0000-0002-0617-0524}}
\affiliation{Van Swinderen Institute of Particle Physics and Gravity, University of Groningen, Groningen 9712 CP, Netherlands}
\affiliation{Department of Physical and Theoretical Chemistry, Faculty of Natural Sciences, Comenius University, Bratislava, Slovakia}

\author{H. A. Perrett}
\affiliation{Department of Physics and Astronomy, The University of Manchester, Manchester M13 9PL, United Kingdom}

\author{{A.~Raggio}\orcidlink{0000-0002-5365-1494}}
\affiliation{Department of Physics, University of Jyv\"{a}skyl\"{a}, Jyv\"{a}skyl\"{a} FI-40014, Finland}

\author{J.~R.~Reilly}
\affiliation{Department of Physics and Astronomy, The University of Manchester, Manchester M13 9PL, United Kingdom}

\author{S.~Rothe\orcidlink{0000-0001-5727-7754}}
\affiliation{Systems Department, CERN, CH-1211 Geneva 23, Switzerland}

\author{{E.~Smets}}
\affiliation{KU Leuven, Instituut voor Kern- en Stralingsfysica, B-3001 Leuven, Belgium}

\author{S.-M. Udrescu}
\affiliation{Department of Physics, Massachusetts Institute of Technology, Cambridge, MA 02139, USA}

\author{B. van den Borne\orcidlink{0000-0003-3348-7276}}
\affiliation{KU Leuven, Instituut voor Kern- en Stralingsfysica, B-3001 Leuven, Belgium}

\author{Q. Wang}
\affiliation{School of Nuclear Science and Technology, Lanzhou University, Lanzhou 730000, China}

\author{{J. Warbinek}}
\affiliation{GSI Helmholtzzentrum f\"ur Schwerionenforschung GmbH, 64291 Darmstadt, Germany}
\affiliation{Department of Chemistry - TRIGA Site, Johannes Gutenberg-Universit\"at Mainz, 55128 Mainz, Germany}

\author{J. Wessolek\orcidlink{0000-0001-9804-5538}}
\affiliation{Department of Physics and Astronomy, The University of Manchester, Manchester M13 9PL, United Kingdom}
\affiliation{Systems Department, CERN, CH-1211 Geneva 23, Switzerland}

\author{X. F. Yang}
\affiliation{School of Physics and State Key Laboratory of Nuclear Physics and Technology, Peking University, Beijing 100971, China}

\author{C. Z\"{u}lch}
\affiliation{Fachbereich Chemie, Philipps-Universit\"{a}t Marburg, Marburg 35032, Germany}

\author{the ISOLDE Collaboration}

\date{\today}

\maketitle

\section{Extended methods}\label{appsec:extended_methods}

\subsection{Data analysis} \label{appsubsec:data_analysis}
The measured wavenumbers in the acquired spectra were firstly Doppler-shifted to the molecular rest frame wavenumber $\Tilde{\nu}$ according to the expression $\Tilde{\nu} = \frac{1-\beta}{\sqrt{1-\beta^2}}\Tilde{\nu}_0$, where $\beta=v/c$ with $c$ the speed of light, and $\Tilde{\nu}_0$ the wavenumber in the lab frame. The velocity of the beam was determined from the ion kinetic energy, which was defined by the platform voltage of the radiofrequency cooler-buncher that drifted over time between 39,905 and 39,910~{\color{black}V}. Fluctuations and drifts of the platform voltage were monitored by continuous measurements (1 measurement per second) of the real voltage using a calibrated potential divider (PTB PT60-2) and a digital multimeter (Agilent 34401A). The voltage measurements (precision of 10~mV at 40~kV) were then used to accurately determine the velocity of the $^{226}$RaF beam for each wavenumber measurement. Following Doppler-shifting, the spectra {\color{black}obtained with fundamental Ti:Sa ($\sim$3~GHz) and the narrower ($\sim$0.8~GHz) Cobra dye laser} were binned {\color{black}into a histogram} with a bin size of 3~GHz, {\color{black}and the spectra obtained with the broader ($\sim$9~GHz) Spectra Physics dye laser and second-harmonic Ti:Sa ($\sim$6~GHz) with a bin size of 9~GHz}.

The binned spectra were analyzed using the contour-fitting routine of the {\sc{pgopher}} package {\color{black}using the effective Hamiltonian for linear molecules embedded in {\sc{pgopher}}} ~\cite{Western2017}; for $^2\Sigma$ states, the molecular constants relevant to Hund's case \textit{b} were used, and for $^2\Pi$ and $^2\Delta$ states, the constants for Hund's case \textit{a} were used. The state origin $T_0$, rotational constant~$B$, spin-rotation coupling constant $\gamma$ (only for $^2\Sigma$ states), and $\Lambda$-doubling constant $p$ (only for $^2\Pi$ states) were varied during the fitting routine to reach agreement between simulation and experiment.

All measured spectra involved electronic transitions starting from either the $X$~$^2\Sigma_{1/2}$ or the $A$~$^2\Pi_{1/2}$ states. The molecular parameters of these two states are known from the {\color{black}rotationally resolved ($\Delta f \approx 100$~MHz)} laser spectroscopy of the $A$~$^2\Pi_{1/2} \leftarrow X$~$^2\Sigma_{1/2}$ transition in $^{226}$RaF~\cite{Udrescu2023}. As a result, in the present study only the properties of the upper vibronic state in each spectrum were varied during the fitting procedure.

The statistical uncertainty in the excitation energies was extracted by the standard deviation of the fitted parameter given by the contour-fitting routine of {\sc{pgopher}}. The uncertainty in the raw data was the error in the count rate (y-axis) in the spectra, which was determined as the square root of the number of data points in each bin. The standard deviations of the fitted excitation energies were scaled by the square root of the reduced chi-squared of the fit $\sqrt{\chi_r^2}$. {\color{black}These errors are denoted in parentheses in both the main text and this supplemental material.}

A systematic uncertainty is considered for all excitation energies, which corresponds to the Voigt-peak linewidth set in {\sc{pgopher}} to best match the observed linewidths for each spectrum, and aims to account for the propensity of the contour-fitting routine to converge to local minima, as it cannot move a simulated line by more than the set linewidth~\cite{Western2017}. An additional component of 0.02~cm$^{-1}$ is added to account for the combined sources of systematic error stemming from the experimental equipment ({\color{black}the sources of this error are discussed in} Ref.~\cite{Udrescu2023}). {\color{black}These errors are denoted in square brackets in both the main text and this supplemental material.}

The best-fit values and uncertainties of the excitation energies were also obtained independently using a chi-squared minimization code written in Python and the correlated errors were determined by inspecting the corner plots of the fitted parameters. The results of the independent fitting were consistent with the results from the {\sc{pgopher}} analysis, and thus the latter are used here.

The resolution of the spectra of transitions to the $G$~$^{2}\Pi_{1/2}$ and $E$~$^{2}\Sigma_{1/2}$ states was high enough to enable an analysis using the line-fitting routine of {\sc{pgopher}}, which was found to yield results in agreement within 1$\sigma$ with contour-fitting of the same spectra. Only contour-fitting results are included in this work for consistency.

\subsection{Calculations}
In the large-scale calculations described in \textit{Methods}, we calculated differences between total energies of the excited electronic states and the ground state at the internuclear distance 
2.24 \AA, which 
corresponds
to the equilibrium internuclear distance of the electronic ground state of RaF. However, to account for the difference in total energies 
at the equilibrium distance of each state, it was necessary to determine the equilibrium internuclear distances for all states 
(see Table~\ref{tab6}). The obtained ``non-verticality'' contributions were added to the vertical ``bulk'' 69-electron FS-RCCSD excitation energies obtained in the large-scale calculations. To calculate these contributions, potential energy curves for all considered electronic states were calculated.

Similarly to Ref.~\cite{Zaitsevskii2022}, the potential energy curve for the electronic ground state was calculated within the two-component (valence) GRECP approach using the single-reference coupled cluster with single, double and perturbative triple cluster amplitudes CCSD(T) method. To obtain potential energy curves for excited electronic states as a function of the internuclear distance $R$,  the excitation energy calculated at the FS-RCCSD level for a given value of $R$ was added to the energy of the electronic ground state for the same value of $R$. In these calculations, 37 electrons were included in the correlation treatment. For Ra, we used the $[25s\, 17p\, 12d\, 9f\, 4g\, 2h]$ basis set optimized for the GRECP calculations, while for F the uncontracted Dyall's AAEQZ basis set~\cite{Dyall:2016} was employed.

To characterize molecular terms in the $\Lambda S$ coupling scheme, the mean values of the electron orbital angular momentum projection operator on the molecular axis were calculated at the FS-RCCSD level (within the finite-field approach) and rounded up to an integer using the code to compute corresponding matrix elements developed in Ref.~\cite{Skripnikov2016}.

In the uncertainty estimations, it is considered that the basis set correction (CBS) has an uncertainty of 50\%. The calculations did not take into account the retardation part of the Breit interaction. According to the data for the Ra$^+$ ion~\cite{Skripnikov2021,Eliav:1996}, the contribution of the retardation part of the Breit interaction to the transition energies in Ra$^+$ is within 20\% of the Gaunt interaction effect. Thus, the uncertainty due to the excluded retardation part of the Breit interaction is estimated as 20\% of the Gaunt-effect contribution. The accuracy of the model QED operator approach is high~\cite{Shabaev:13,Kaygorodov:2019}. Therefore, the uncertainty of the QED correction is suggested to be in the order of 20\%. The uncertainty of the high-order correlation effects (beyond the FS-RCCSDT model) can be estimated by comparing the FS-RCCSDT result with the single-reference coupled cluster with single, double, triple and perturbative quadruple cluster amplitudes, CCSDT(Q)~\cite{Kallay:6,MRCC2020}. According to our calculations performed for states of RaF with near single-reference character, the correlation corrections to FS-RCCSD calculated with the FS-RCCSDT and CCSDT(Q) methods agree within 60\%. To be more conservative, in the uncertainty estimation 
we set the uncertainty of the higher-order correlation contribution to 75\%. 
The final uncertainty estimation of the theoretical electronic excitation energies was calculated as the square root of the sum of squares of the uncertainties described above for each state.

\subsection{Spin-orbit constants} \label{appsubsec:so}
The molecular spin-orbit (SO) constants $A$ were extracted as:
\begin{equation}
    A = \frac{1}{2S\Lambda} \big( E_{\Lambda+S} - E_{\Lambda-S} \big)
\end{equation}
All electronic states in alkaline-earth monofluorides have a single unpaired valence electron, and thus $S=1/2$. For $^2\Pi$ states, $\Lambda = 1$, while for $^2\Delta$ states, $\Lambda = 2$. For the SO constants in RaF, the observed excitation energies reported in Table~1 of the main text were used. 

For the microscopic SO constants $\zeta$ in 
Ra$^+$, the excitation energies in the NIST Atomic Spectra Database were used. For $^2P$ states:
\begin{equation}
    \zeta = \frac{2}{3} \big( E_{L+S} - E_{L-S} \big)
\end{equation}
and for $^2D$ states:
\begin{equation}
    \zeta = \frac{2}{5} \big( E_{L+S} - E_{L-S} \big)
\end{equation}

\subsection{State assignment} \label{appsubsec:assignment}
The following subsections provide details that guided the assignment of term symbols for the observed spectra. In addition to the specifics of each spectrum, the assignment was guided by the agreement between the experimentally determined rotational constant of the $v=0$ state $B_{\rm{0,obs}}$ and the theoretically determined equilibrium rotational constant $B_{\rm{e,th}}$ across all states. While $B_{\rm{0,obs}}$ and $B_{\rm{e,th}}$ are different molecular constants, they are closely related and at the given level of precision a large deviation between the two would indicate an incorrect assignment.

\subsubsection{$A$~$^2\Pi_{3/2}$}
The spectrum of the transition from the electronic ground state to $A$~$^2\Pi_{3/2}$ was measured in the region where it had been previously measured in Ref.~\cite{GarciaRuiz2020}. The spectrum was {\color{black}analyzed in this work using {\sc pgopher} and the excitation energy was extracted from the molecular Hamiltonian, leading to a small correction and re-assignment compared to the previous result reported in Ref.~\cite{GarciaRuiz2020}. The measured spectrum was congested, with the contours of the vibrational transitions being largely overlapped, which could explain the deviation of the observed vibrational spacing $\omega_{\rm{obs}}$ from $\omega_{\rm{th}}$ in Table~\ref{tab6}.}

\begin{table*}[h]
\begin{minipage}{440pt}
\caption{{Comparison of electronic excitation wavenumbers ($T_0$, in~cm$^{-1}$) assigned to low-lying states in RaF between Ref.~\cite{GarciaRuiz2020} and the present work.} All assignments refer to the $v=0$ vibrational state of each electronic state.}\label{tabapp:newassign}%
\begin{tabular}{cccccc}
\hline \hline
                             & Garcia et al. 2020 & This work & Theory \\ \hline
$B$~$^2\Delta_{3/2}$ & 15,142.7(5)*    & 14,333.00(161)[51] & 14,300(61) \\
($B$~$^2\Delta_{5/2}$) & ---             & 15,140.36(48)[51] &       15,099(70)\\
$A$~$^2\Pi_{3/2}$    & 15,344.6(50)    & 15,335.73(49)[62] & 15,355(35)\\
$C$~$^2\Sigma_{1/2}$ & 16,175.2(5)$^\$$     & 16,612.06(18)[51] & 16,615(69)
\\\hline\hline
\end{tabular}
\begin{flushleft}
* Tentatively assigned in Ref.~\cite{GarciaRuiz2020}.\\
$^\$$ Following the new measurements, this transition is re-interpreted as $C$~$^2\Sigma_{1/2}\leftarrow$ $X$~$^2\Sigma_{1/2}$ $(v'=0\leftarrow v''=1)$. We place the $v'=0\leftarrow v''=0$ transition at 16,612.06(18)[51]~cm$^{-1}$, in agreement with the theoretical prediction.
\end{flushleft}
\end{minipage}
\end{table*}

\subsubsection{$B$~$^2\Delta_{3/2}$}
{\color{black}The excitation energy of the $B$~$^2\Delta_{3/2}$ state was assigned via the spectrum of a transition from the $X$~$^2 \Sigma_{1/2}$ electronic ground state, where a vibrational progression was observed and interpreted as belonging to the diagonal transitions $\Delta v=0$ with $v''=0-3$. The observed excitation energy of 14,333.00(161)[51]~cm$^{-1}$ and the observed vibrational spacing of $\omega_{\rm{obs}}=425.9(34)$~cm$^{-1}$ are in excellent agreement with the calculated values of 14,300(61)~cm$^{-1}$ and $\omega_{\rm{th}}=426(4)$~cm$^{-1}$, respectively, for the $B$~$^2\Delta_{3/2}$ state, while the theoretical values disagree with the previous tentative assignment in Ref.~\cite{GarciaRuiz2020}. As a result, the new assignment for the excitation energy of $B$~$^2\Delta_{3/2}$ is adopted instead of the previous tentative assignment at 15,142.7(5)~cm$^{-1}$ from Ref.~\cite{GarciaRuiz2020}.
}

\begin{figure*}
    \centering
    \includegraphics[width=0.95\textwidth]{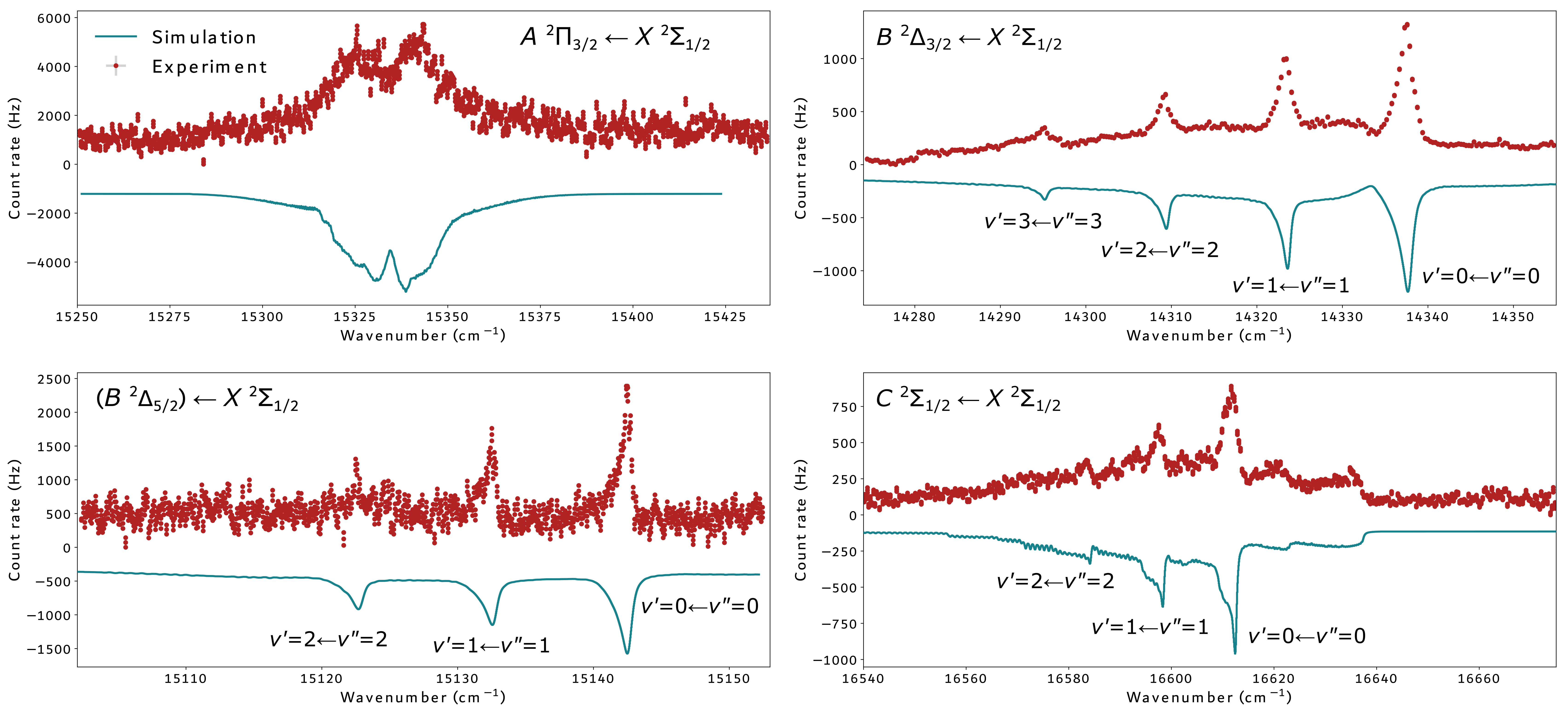}
    \caption{Measured and simulated spectra for the transitions from the ground state to the low-lying states. The different vibrational bands are overlapped in the spectrum to $A$~$^2\Pi_{3/2}$, but are included in the simulation.}
    \label{fig:SM_allspectra_1}
\end{figure*}

\subsubsection{$B$~$^2\Delta_{5/2}$}
The previously measured~\cite{GarciaRuiz2020} spectrum that was tentatively assigned as a transition to $B$~$^2\Delta_{3/2}$ is now tentatively reassigned as a transition to $B$~$^2\Delta_{5/2}$ {\color{black}and analyzed using {\sc pgopher}, extracting both the excitation energy and the vibrational spacing}.

{\color{black}The observed excitation energy is in agreement with the theoretical prediction for the $B$~$^2\Delta_{5/2}$ state and the observed vibrational spacing of $\omega_{\rm{obs}}= 428.6(11)$~cm$^{-1}$ is also in good agreement with $\omega_{\rm{th}}=430(4)$~cm$^{-1}$ (Table~\ref{tab6}).

The transition $B$~$^2\Delta_{5/2} \leftarrow X$~$^2\Sigma_{1/2}$ is expected to be dipole-forbidden {\color{black}(due to $\Delta \Lambda = 2$ and $\Delta \Omega = 2$)} but allowed by the $L$-uncoupling interaction~\cite{LefebvreBrionField2004} of the $B$~$^2\Delta_{5/2}$ state with $A$~$^2\Pi_{3/2}$. In the absence of additional information that can be used to unambiguously  identify the upper state, this assignment remains tentative.

\subsubsection{$C$~$^2 \Sigma_{1/2}$}
The new assignment for the energy of the $C$~$^2 \Sigma_{1/2}$ state follows from the observation of a transition from the electronic ground state {\color{black} at 16,612.06(18)[51]~cm$^{-1}$}, in close agreement with the prediction at 16,615(69)~cm$^{-1}$. This transition lies outside the range that was scanned in Ref.~\cite{GarciaRuiz2020} and was thus not previously observed. The spectral profile of this transition is very similar to that of the transition at 16,175.2~cm$^{-1}$ that was previously assigned as $C$~$^2\Sigma_{1/2} \leftarrow X$~$^2\Sigma_{1/2}$ ($v'=0 \leftarrow v''=0$), but its intensity is twice that of the transition at  16,175.2~cm$^{-1}$. As a result, it is concluded that it is the newly discovered transition that in fact corresponds to $C$~$^2\Sigma_{1/2} \leftarrow X$~$^2\Sigma_{1/2}$ ($v'=0 \leftarrow v''=0$) {\color{black}and the previously identified transition corresponds to ($v'=0 \leftarrow v''=1$)}.


{\color{black}\subsubsection{$D$~$^2\Pi_{1/2}$ and $D$~$^2\Pi_{3/2}$}
The excitation energies of the $D$~$^2\Pi_{1/2}$ and $D$~$^2\Pi_{3/2}$ states were assigned based on transitions starting from the electronic ground state $X$~$^2\Sigma_{1/2}$.

For each state, the diagonal vibrational progression of the transition from the ground state was observed up to $v'=2 \leftarrow v''=2$. This allowed the extraction of the vibrational spacing for the two states, giving an observed value of $\omega_{\rm{obs}}=431.1(4)$~cm$^{-1}$ for $D$~$^2\Pi_{1/2}$ as opposed to the calculated value of $\omega_{\rm{th}}=434(4)$~cm$^{-1}$, and $\omega_{\rm{obs}}=441.8(8)$~cm$^{-1}$ for $D$~$^2\Pi_{3/2}$ as opposed to $\omega_{\rm{th}}=444(4)$~cm$^{-1}$ (Table~\ref{tab6}). The agreement with theory in terms of magnitude as well as observing $\omega$ to be larger for the $D$~$^2\Pi_{3/2}$ state (and larger than for the ground state) support the assignment in this work. The contours of the spectra interpreted as corresponding to transitions to the $D$~$^2\Pi_{1/2}$ vibrational states are distinctly different from the spectra interpreted as corresponding to the $D$~$^2\Pi_{3/2}$ states. Therefore, the possibility of the two sets of spectra corresponding to one diagonal and one non-diagonal vibrational progressions ($\Delta v=0$ and $\Delta v = \pm 1$) to the same upper electronic state is excluded.

The spectrum of the transition to the $D$~$^2\Pi_{1/2}$ state was highly congested with the contours of the vibrational transitions largely overlapping, so a contribution from $v'=3 \leftarrow v''=3$ was also taken into account as its presence could not be excluded. This contribution affected the results of the $v'=0$ state within uncertainties.

As a validity check for the assignment of $D$~$^2\Pi_{1/2}$ without the use of the theoretical calculations, the $\Lambda$-doubling constant $p$ extracted from the contour fit (see Table~\ref{tab6}) can be compared to the value extracted via the pure precession approximation for the interaction between the $^2\Pi$ state and neighboring $^2\Sigma$ states~\cite{Mulliken1931}:
\begin{equation}\label{eq:p_pure}
    p_{\rm{pure}} = 2AB_v l(l+1) / \Delta E_{\Pi\Sigma}
\end{equation}

Using $l=1$, $B=0.1922(4)$ (Table~\ref{tab6}), and $A=362(1)$, and considering interaction only with the neighboring $C$~$^2\Sigma_{1/2}$ state, the approximation results in $p_{\rm{pure}}=0.049$. This value is in agreement with the value extracted from the contour fit (Table~\ref{tab6}).

\begin{figure*}
    \centering
    \includegraphics[width=0.95\textwidth]{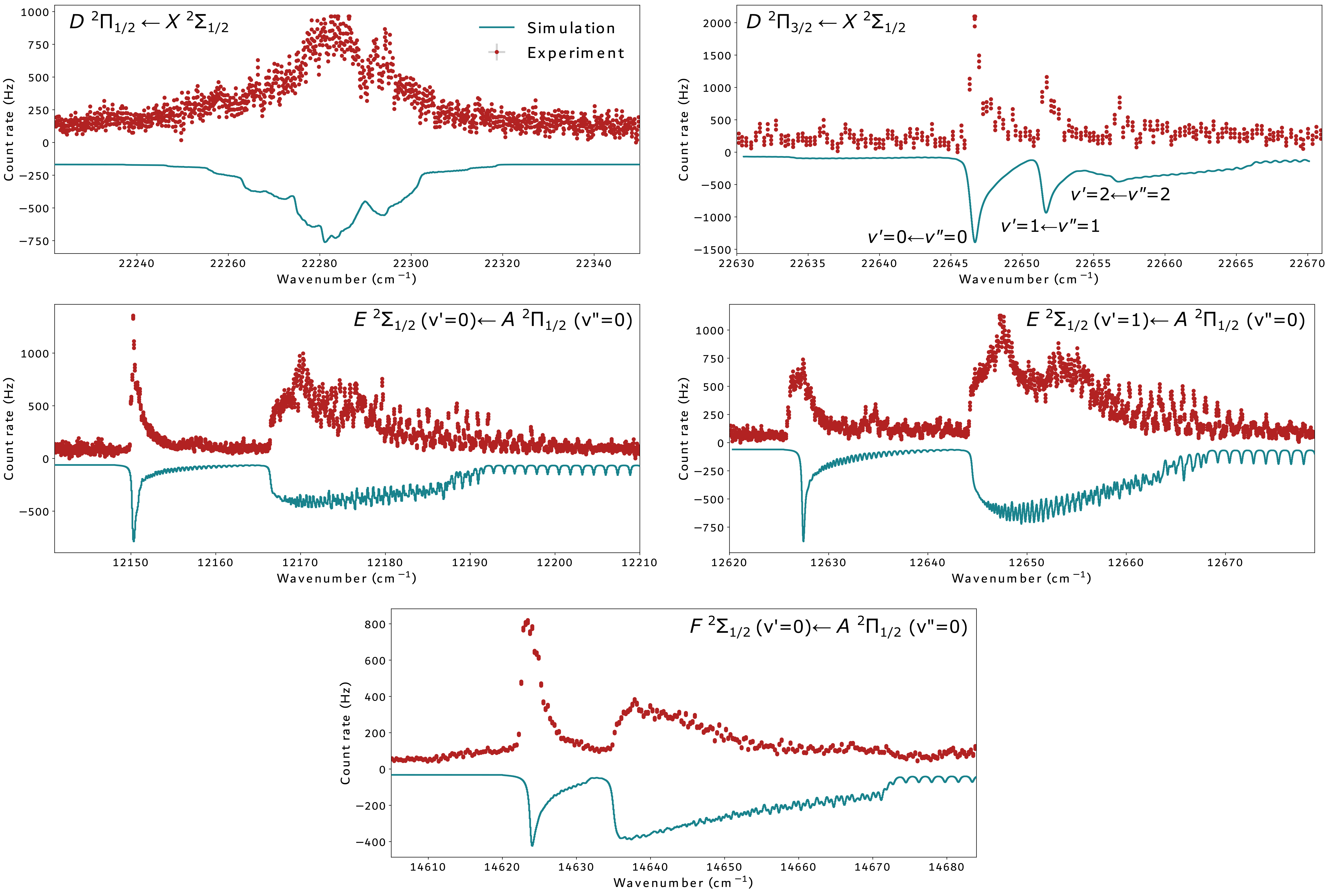}
    \caption{Measured and simulated spectra for the transitions from the ground state to the low-lying states. The different vibrational bands are overlapped in the spectrum of $D$~$^2\Pi_{1/2} \leftarrow A$~$^2\Pi_{1/2}$, but are included in the simulation.}
    \label{fig:SM_allspectra_2}
\end{figure*}

\subsubsection{$E$~$^2\Sigma_{1/2}$}
The energy of the $E$~$^2\Sigma_{1/2}$ state was assigned based on a transition starting from $A$~$^2\Pi_{1/2}$ ($v=0$), as the first laser step was exciting the $A$~$^2\Pi_{1/2}$ ($v'=0$) $\leftarrow$ $X$~$^2\Sigma_{1/2}$ ($v''=0$) transition.

Only two transitions were discovered in a range of 1,500~cm$^{-1}$ around the predicted excitation energy of the $E$~$^2\Sigma_{1/2}$ state. The spectra of both transitions have very similar shapes and are separated by $\omega_{\rm{obs}}=478.0(2)$~cm$^{-1}$, which is in close agreement with $\omega_{\rm{th}}=481(5)$~cm$^{-1}$ (Table~\ref{tab6}). Since the transition starts from $v=0$ of the lower state, the two measured spectra were interpreted as belonging to transitions to $v=0$ and $v=1$ of the $E$~$^2 \Sigma_{1/2}$ state, which is the only state predicted to lie within a few thousand~cm$^{-1}$ of the measured structures.

It was observed that, for this spectrum in particular, using a Gaussian temperature distribution centered at room temperature improved the reduced chi-squared of the fit.

{\color{black}\subsubsection{$F$~$^2\Sigma_{1/2}$}
The energy of the $F$~$^2\Sigma_{1/2}$ state was assigned based on a transition starting from $A$~$^2\Pi_{1/2}$ ($v=0$), as the first laser step was exciting the $A$~$^2\Pi_{1/2}$ ($v'=0$) $\leftarrow$ $X$~$^2\Sigma_{1/2}$ ($v''=0$) transition. Therefore, the spectrum was interpreted as $F$~$^2\Sigma_{1/2} \leftarrow A$~$^2 \Pi_{1/2}$ ($v'=0\leftarrow v''=0$). A second spectrum was obtained starting from $A$~$^2\Pi_{1/2}$ ($v=1$), interpreted as $F$~$^2\Sigma_{1/2} \leftarrow A$~$^2 \Pi_{1/2}$ ($v'=1\leftarrow v''=1$).

Only one transition was discovered in a range of 500~cm$^{-1}$ around the predicted excitation energy of the $F$~$^2\Sigma_{1/2}$ state, starting from the $A$~$^2\Pi_{1/2}$ $v=0$ state. The contour of the spectrum resembled closely that of the transition to $E$~$^2\Sigma_{1/2}$ $v=0$ indicating that the upper state has a similar wavefunction, as it is expected for the $F$~$^2\Sigma_{1/2}$ state. The high intensity of the measured signal excludes the possibility of the spectrum corresponding to a high overtone transition to a vibrational state of $E$~$^2\Sigma_{1/2}$ (such as $v'=5\leftarrow v''=0$). Therefore, this spectrum is attributed to $F$~$^2\Sigma_{1/2}$ $v=0$ as the upper state and the excitation energy is assigned accordingly.

The observed vibrational spacing $\omega_{\rm{obs}}=491.7(18)$~cm$^{-1}$, extracted using the $0\leftarrow 0$ and $1 \leftarrow 1$ transition frequencies is in close agreement with the calculated $\omega_{\rm{th}}=488(5)$~cm$^{-1}$ (Table~\ref{tab6}).
}

\subsubsection{$G$~$^2\Pi_{1/2}$}
The energy of the $G$~$^2\Pi_{1/2}$ state was assigned based on a transition starting from $A$~$^2\Pi_{1/2}$ ($v=0$), as the first laser step was exciting the $A$~$^2\Pi_{1/2}$ ($v'=0$) $\leftarrow$ $X$~$^2\Sigma_{1/2}$ ($v''=0$) transition.

\begin{figure*}
    \centering
    \includegraphics[width=0.95\textwidth]{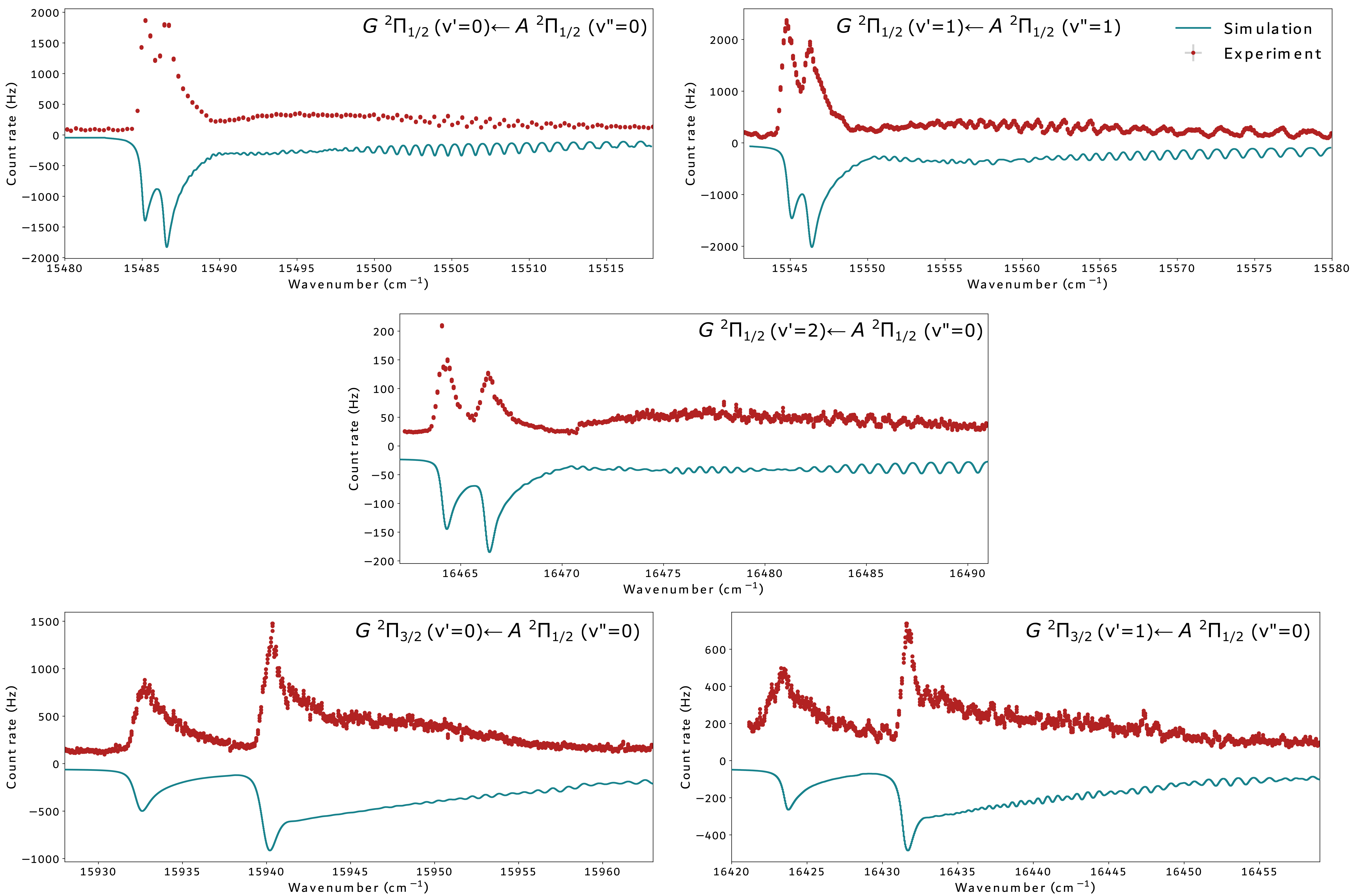}
    \caption{Measured and simulated spectra of diagonal and off-diagonal vibronic transitions from $A$~$^2\Pi_{1/2}$ to the $G$~$^2\Pi$ states.}
    \label{fig:SM_allspectra_3}
\end{figure*}

A spectrum that starts from the same lower state as the one assigned as $G$~$^2\Pi_{1/2} \leftarrow A$~$^2\Pi_{1/2}$ ($v'=0 \leftarrow v''=0$) and has a very similar spectral profile was measured at approximately 980~cm$^{-1}$ higher in energy (corresponding to approximately two times the calculated vibrational spacing for this state, Table~\ref{tab6}), which was interpreted as a transition to $G$~$^2\Pi_{1/2}$ ($v=2$). Additionally, in a transition starting from the $A$~$^2\Pi_{1/2}$ ($v=1$) state, a spectrum was observed whose upper state was at an excitation energy half-way between those of the $v=0$ and $v=2$ states, with a very similar spectral profile. Therefore, this third spectrum was interpreted as belonging to a transition to $G$~$^2\Pi_{1/2}$ ($v=1$). 
The only other predicted state in the vicinity of $G$~$^2\Pi_{1/2}$ is $G$~$^2\Pi_{3/2}$. The transition strength of $G$~$^2\Pi_{1/2} \leftarrow A$~$^2\Pi_{1/2}$ ($v'=0 \leftarrow v''=0$) was the largest among all transitions measured from $A$~$^2\Pi_{1/2}$ {\color{black}in this experiment}, which is highly improbable for the nominally forbidden transition $G$~$^2\Pi_{3/2} \leftarrow A$~$^2\Pi_{1/2}$. Therefore, the observed spectrum was assigned as $G$~$^2\Pi_{1/2} \leftarrow A$~$^2\Pi_{1/2}$ ($v'=0 \leftarrow v''=0$).

Using the pure precession approximation, as done for $D$~$^2\Pi_{1/2}$, and assuming mixing only with the neighboring ($H$~$^2\Sigma_{1/2}$) state, $p_{\rm{pure}}=-0.407$ is in good agreement with the fitted value $p_{\rm{obs}}=-0.360(42)$.

\subsubsection{$G$~$^2\Pi_{3/2}$}
The energy of the $G$~$^2\Pi_{3/2}$ state was assigned based on a transition starting from $A$~$^2\Pi_{1/2}$ ($v=0$), as the first laser step was exciting the $A$~$^2\Pi_{1/2}$ ($v'=0$) $\leftarrow$ $X$~$^2\Sigma_{1/2}$ ($v''=0$) transition.

Within a range of 400~cm$^{-1}$ around the prediction for the transition energy to the $G$~$^2\Pi_{3/2}$ state, spectra of two transitions were found, and the excitation energy of both would be in agreement with the prediction.

However, for only one of the two transitions, a second spectrum also starting from $A$~$^2\Pi_{1/2}$ ($v=0$) was measured at 491.6(4)~cm$^{-1}$ higher that has a very similar spectral profile. This wavenumber difference is in agreement with the calculated vibrational spacing for the $G$~$^2\Pi_{3/2}$ state (Table~\ref{tab6}), and so these two spectra that had the same spectral profile were interpreted as transitions to $v=0$ and $v=1$ of $G$~$^2\Pi_{3/2}$.

\subsubsection{$H$~$^2\Sigma_{1/2}$, $I$~$^2 \Delta_{3/2}$, and $I$~$^2\Delta_{5/2}$}
The energies of these three states were assigned based on transitions starting from $A$~$^2\Pi_{1/2}$ ($v=0$), as the first laser step was exciting the $A$~$^2\Pi_{1/2}$ ($v'=0$) $\leftarrow$ $X$~$^2\Sigma_{1/2}$ ($v''=0$) transition in all cases.

\begin{figure}
    \centering
    \includegraphics[width=0.45\textwidth]{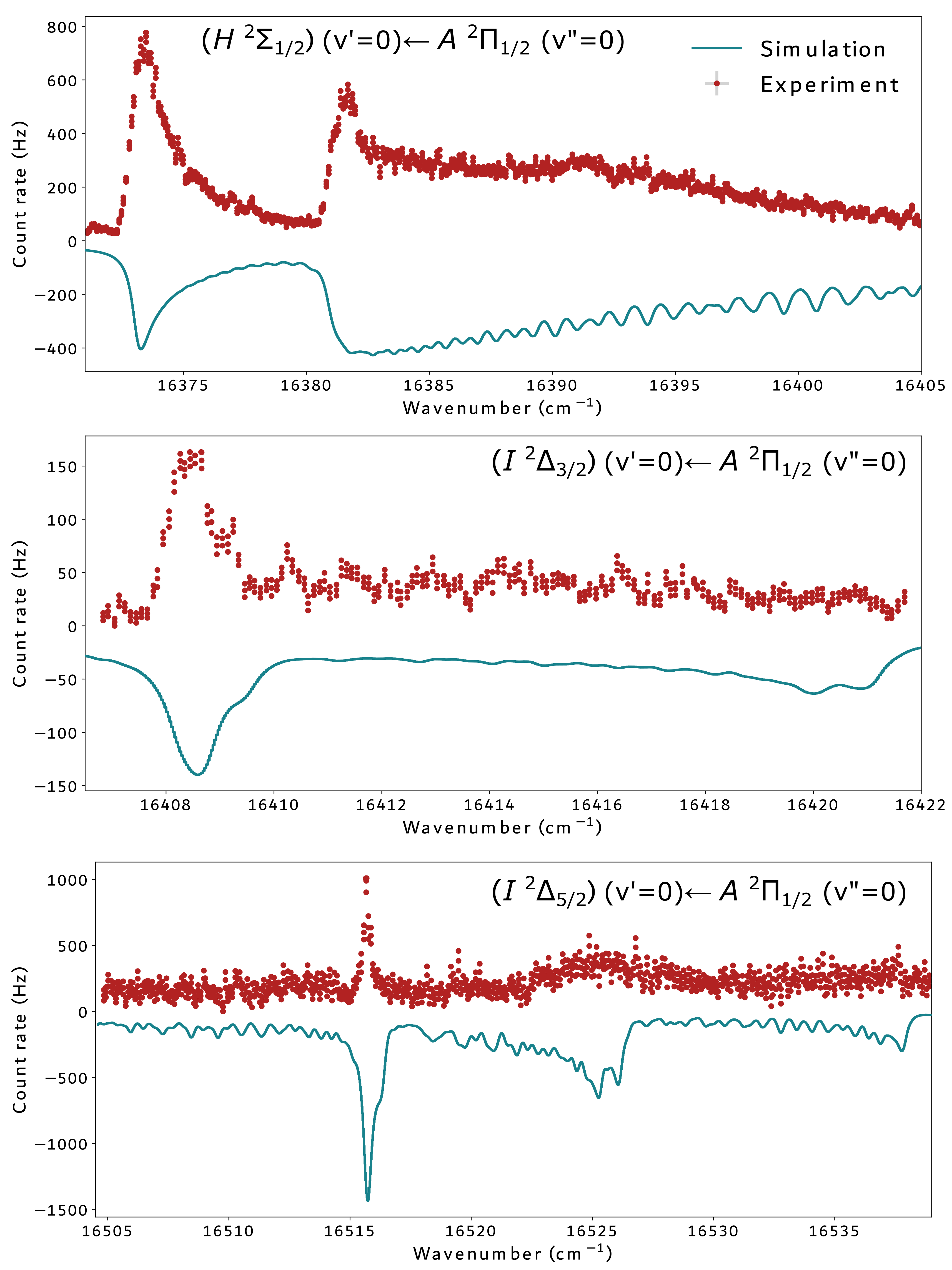}
    \caption{Measured and simulated spectra from $A$~$^2\Pi_{1/2}$ to the tentatively assigned highest-lying states in this study.}
    \label{fig:SM_allspectra_4}
\end{figure}

In the vicinity of the transitions assigned to these three states, multiple transitions were observed. Two of those transitions were identified as the $v'=1 \leftarrow v''=0$ and $v'=2 \leftarrow v''=0$ transitions to the $G$~$^2\Pi_{3/2}$ and $G$~$^2\Pi_{1/2}$ states, respectively.

At approximately 16,175~cm$^{-1}$, the transition $C$~$^2\Sigma_{1/2} \leftarrow X$~$^2\Sigma_{1/2}$ ($v'=0 \leftarrow v''=1$) was identified, while at approximately 16,610~cm$^{-1}$, the transition $C$~$^2\Sigma_{1/2} \leftarrow X$~$^2\Sigma_{1/2}$ ($v'=0 \leftarrow v''=0$) was identified. These two transitions did not arise from the combined effect of all three lasers in the excitation scheme, but only from the scanning laser and the non-resonant ionization step. Therefore, they were identified as transitions from the electronic ground state, rather than from $A$~$^2\Pi_{1/2}$ ($v=0$).

Two bands were assigned as transitions to $H$~$^2\Sigma_{1/2}$ ($v=0$) and $I$~$^2\Delta_{3/2}$ ($v=0$) based on the computational predictions. As the predicted energies of the two states are within uncertainties of each other, their assignment is tentative.

The spectrum that was assigned as belonging to the transition to $I$~$^2\Delta_{5/2}$ was identified due to its significantly lower intensity compared to all other spectra, which is consistent with the dipole-forbidden $I$~$^2\Delta_{5/2} \leftarrow A$~$^2 \Pi_{1/2}$ ($v'=0 \leftarrow v''=0$) transition. As the transition is dipole-forbidden {\color{black}due to $\Delta \Omega = 2$}, the assignment is tentative.

One more, very weak spectral feature was identified at an excitation energy of approximately 29,650~cm$^{-1}$, whose spectral profile suggests the form of a $^2\Sigma \leftarrow ^2\Pi$ spectrum. Neither a firm nor a tentative assignment was possible at the present time. Based on the $T_0$ and $\omega_{\rm{obs}}$ for the $F$~$^2 \Sigma_{1/2}$ state, this spectrum could correspond to {\color{black}an overtone} transition {\color{black}of} $F$~$^2 \Sigma_{1/2} \leftarrow A$~$^2 \Pi_{1/2}$ ({\color{black}such as} $v'=3 \leftarrow v''=0$), whose expected strength is consistent with the very low intensity of the observed structure. However, {\color{black}more information on the anharmonicity of the vibrational spacing of $F$~$^2 \Sigma_{1/2}$ for higher values of $v$ is required to make this assignment definite.} 

\begin{table*}[ht]
\begin{minipage}{440pt}
\caption{{Theoretical 
adiabatic electronic excitation energies ($T_e$, in cm$^{-1}$) calculated at the 69e-FS-RCCSD-extAE4Z level including higher-order corrections.} Each column presents cumulative results, adding a contribution to the values in the column to its left. In the final column, the zero-point vibrational energy is added to $T_e$ to arrive at $T_0$ that is compared with experiment. }\label{tab3}%
\begin{tabular}{ccccccc}
\hline \hline
State              & 69e-extAE4Z & +27e-CCSDT & +CBS & +Gaunt & +QED & +ZPE ($T_0$) \\\hline
X $^2\Sigma_{1/2}$ & 0           & 0           & 0     & 0       & 0   & 0  \\
A $^2\Pi_{1/2}$    & 13,406       & 13,360       & 13,354 & 13,359   & 13,301 &  13,299\\
B $^2\Delta_{3/2}$ & 14,582       & 14,531       & 14,444 & 14,379   & 14,307 &  14,300\\
B $^2\Delta_{5/2}$ & 15,403       & 15,336       & 15,249 & 15,171   & 15,103 &  15,099\\
A $^2\Pi_{3/2}$    & 15,481       & 15,438       & 15,432 & 15,414   & 15,357 &  15,355\\
C $^2\Sigma_{1/2}$ & 16,785       & 16,694       & 16,685 & 16,674   & 16,622 &  16,615\\
D $^2\Pi_{1/2}$    & 22,724       & 22,504       & 22,442 & 22,388   & 22,323 &  22,320\\
D $^2\Pi_{3/2}$    & 23,074       & 22,852       & 22,790 & 22,734   & 22,671 &  22,673\\
E $^2\Sigma_{1/2}$ & 25,616       & 25,509       & 25,554 & 25,544   & 25,501 &  25,520\\
F $^2\Sigma_{1/2}$ & 28,410       & 28,067       & 28,080 & 28,050   & 27,995 &  28,019\\
G $^2\Pi_{1/2}$    & 28,962       & 28,818       & 28,865 & 28,850   & 28,797 &  28,824\\
G $^2\Pi_{3/2}$    & 29,391       & 29,276       & 29,323 & 29,309   & 29,257 &  29,284\\
H $^2\Sigma_{1/2}$ & 29,860       & 29,654       & 29,697 & 29,681   & 29,629 &  29,663\\
I $^2\Delta_{3/2}$ & 29,862       & 29,729       & 29,765 & 29,741   & 29,686 &  29,715\\
I $^2\Delta_{5/2}$ & 30,006       & 29,868       & 29,904 & 29,878   & 29,824 &  29,852
\\\hline\hline
\end{tabular}
\end{minipage}
\end{table*}

\begin{table*}[ht]
\begin{minipage}{440pt}
\caption{{Theoretical results (in cm$^{-1}$) calculated at the FS-RCCSD-extAE3Z level for different numbers of correlated electrons.} Unless specified in parenthesis, the shell contributions refer to electron shells belonging to the Ra atom.}\label{tab4}%
\begin{tabular}{cccccccccc}
\hline\hline
\multicolumn{1}{c}{\multirow{4}{*}{State}} & \multirow{4}{*}{97e}& \multirow{4}{*}{69e} & \multirow{4}{*}{35e} & \multirow{4}{*}{27e} & \multirow{4}{*}{17e} & \multicolumn{4}{c}{Shell contributions}     \\
\multicolumn{1}{c}{}&&&&&&& $1s$(F) && \\
\multicolumn{1}{c}{}&&&&&& $1s,...,3d$ & $4s4p4d4f$ & $5s5p$  & $5d$ \\
\multicolumn{1}{c}{}&&&&&& (97e-69e) & (69e-35e) & (35e-27e) & (27e-17e) \\\hline
X $^2\Sigma_{1/2}$& 0  & 0 & 0   & 0  & 0   & 0 & 0 & 0  & 0 \\
A $^2\Pi_{1/2}$   & 13,396& 13,396& 13,356& 13,346& 13,061& 0& 40& 10& 284\\
B $^2\Delta_{3/2}$& 14,604& 14,604& 14,528& 14,491& 14,133& 0& 76& 37& 358\\
B $^2\Delta_{5/2}$& 15,424& 15,423& 15,356& 15,318& 14,920& 0& 67& 38& 397\\
A $^2\Pi_{3/2}$   & 15,476& 15,475& 15,423& 15,400& 15,084& 2& 51& 24& 316\\
C $^2\Sigma_{1/2}$& 16,792& 16,790& 16,745& 16,727& 16,446& 2& 45& 18& 280\\
D $^2\Pi_{1/2}$   & 22,899& 22,899& 22,837& 22,814& 22,443& 1& 62& 23& 371\\
D $^2\Pi_{3/2}$   & 23,238& 23,237& 23,177& 23,154& 22,769& 1& 60& 23& 385\\
E $^2\Sigma_{1/2}$& 25,582& 25,582& 25,542& 25,536& 25,208& 0& 40& 7 & 328\\
F $^2\Sigma_{1/2}$& 28,457& 28,457& 28,400& 28,389& 27,995& 0& 57& 11& 394\\
G $^2\Pi_{1/2}$   & 28,985& 28,985& 28,936& 28,926& 28,542& 0& 49& 9 & 384\\
G $^2\Pi_{3/2}$   & 29,400& 29,400& 29,351& 29,339& 28,955& 1& 49& 11& 384\\
H $^2\Sigma_{1/2}$& 30,009& 30,009& 29,960& 29,950& 29,565& 0& 49& 10& 385\\
I $^2\Delta_{3/2}$& 29,926& 29,926& 29,873& 29,861& 29,457& 0& 53& 13& 404\\
I $^2\Delta_{5/2}$& 30,078& 30,078& 30,026& 30,013& 29,605& 0& 52& 13& 408\\
\hline\hline
\end{tabular}
\end{minipage}
\end{table*}

\begin{table*}[ht]
\begin{minipage}{440pt}
\caption{{Theoretical electronic excitation energies (in cm$^{-1}$) calculated  with FS-RCCSD using the CV4Z and augmented CV4Z basis sets, for 27- and 69-electron correlation spaces.} The results obtained with the non-augmented CV4Z basis set could not conclusively identify the energy of the three highest-lying states in this study.}\label{tab5}%
\begin{tabular}{cccc}
\hline\hline
State              & 27e-CV4Z & 69e-CV4Z & 69e-augCV4Z \\ \hline
X $^2\Sigma_{1/2}$ & 0        & 0        & 0           \\
A $^2\Pi_{1/2}$    & 13,517    & 13,542    & 13,425       \\
B $^2\Delta_{3/2}$ & 15,009    & 15,079    & 14,770       \\
B $^2\Delta_{5/2}$ & 15,856    & 15,924    & 15,478       \\
A $^2\Pi_{3/2}$    & 15,648    & 15,690    & 15,549       \\
C $^2\Sigma_{1/2}$ & 17,079    & 17,112    & 16,773       \\
D $^2\Pi_{1/2}$    & 23,491    & 23,537    & 22,762       \\
D $^2\Pi_{3/2}$    & 23,819    & 23,864    & 23,105       \\
E $^2\Sigma_{1/2}$ & 26,052    & 26,069    & 25,739       \\
F $^2\Sigma_{1/2}$ & 29,149    & 29,185    & 28,671       \\
G $^2\Pi_{1/2}$    & 30,779    & 30,804    & 29,152       \\
G $^2\Pi_{3/2}$    & 31,362    & 31,390    & 29,580       \\
H $^2\Sigma_{1/2}$ & ---      & ---      & 30,105       \\
I $^2\Delta_{3/2}$ & ---      & ---      & 30,118       \\
I $^2\Delta_{5/2}$ & ---      & ---      & 30,268      \\
\hline\hline
\end{tabular}
\end{minipage}
\end{table*}

\begin{table*}[ht]
\begin{minipage}{440pt}
\caption{{\color{black}Calculated equilibrium bond length $r_e$, harmonic vibrational spacing $\omega_{\rm{th}}$, and equilibrium rotational constant $B_{\rm{e,th}}$ along with experimentally observed vibrational spacing $\omega_{\rm{obs}}$, rotational constant $B_{\rm{0,obs}}$, spin-rotation coupling constant $\gamma_{\rm{obs}}$, and $\Lambda$-doubling constant $p_{\rm{obs}}$ for the states under 30,000~cm$^{-1}$ from the ground state. All constants are in units of cm$^{-1}$. Information on the X $^2\Sigma_{1/2}$ and A $^2\Pi_{1/2}$ states is from Refs.~\cite{Udrescu2023,GarciaRuiz2020}. Not all known constants are shown for these two states. Details on the assignment and analysis of each state are shown in the subsection titled~\textit{State assignment} in this supplemental material}. The values for $p_{\rm{pure}}$ give the estimates from the pure precession approximation, used as a validity check. Conclusions on the structure of the excited states based on the molecular constants should be cautious of the limited power of contour fitting in determining $B$, $\gamma$, and $p$, and rotationally resolved spectroscopy is necessary to precisely determine the constants.}\label{tab6}%
\begin{tabular}{cccccccccc}
\hline\hline
State              & $r_e$ ($\AA$) & $\omega_{\rm{th}}$ & $\omega_{\rm{obs}}$ & $B_{\rm{e,th}}$ & $B_{\rm{0,obs}}$ & $\gamma_{\rm{obs}}$  & $p_{\rm{obs}}$ &  $p_{\rm{pure}}$\\ \hline
X $^2\Sigma_{1/2}$ & 2.240(10) & 440(4) & 438.4(7) & 0.1917(17) &  0.191985(5)[15] & 0.00585(3)[7] & &\\
A $^2\Pi_{1/2}$    & 2.244(10) & 435(4) & 432.2(7) & 0.1910(17) &  0.191015(5)[15] & & -0.41071(3)[7] & -0.471 \\
B $^2\Delta_{3/2}$ & 2.255(10) & 426(4) & 425.9(34) & 0.1892(17) &  0.1896(14) &\\
B $^2\Delta_{5/2}$ & 2.249(10) & 430(4) & 428.6(11) & 0.1902(17) &  0.1884(12) &\\
A $^2\Pi_{3/2}$    & 2.240(10) & 437(4) & 444.2(9) & 0.1917(17) &  0.1911(6)  &\\
C $^2\Sigma_{1/2}$ & 2.259(10) & 427(4) & 424.4(3) & 0.1885(17) &  0.1886(4) & 0.566(24)  & \\
D $^2\Pi_{1/2}$    & 2.235(10) & 434(4) & 431.1(4) & 0.1926(17) &  0.1922(4) & & 0.041(22) & 0.049 \\
D $^2\Pi_{3/2}$    & 2.226(10) & 444(4) & 441.8(8) & 0.1941(17) &  0.1938(4)  \\
E $^2\Sigma_{1/2}$ & 2.187(10) & 481(5) & 478.0(2) & 0.2011(17) &  0.1996(2) & 0.070(2) & \\
F $^2\Sigma_{1/2}$ & 2.172(10) & 488(5) & 491.7(18) & 0.2039(17) &  0.2053(24) & 0.014(54) & \\
G $^2\Pi_{1/2}$    & 2.180(10) & 495(5) & 496.2(6) & 0.2024(17) &  0.2007(22) & & -0.360(42) & -0.407 \\
G $^2\Pi_{3/2}$    & 2.181(10) & 496(5) & 491.6(4) & 0.2022(17) &  0.2015(6)   \\
H $^2\Sigma_{1/2}$ & 2.176(10) & 510(5) &  & 0.2031(17) &  0.2108(20) & 0.058(46)    \\
I $^2\Delta_{3/2}$ & 2.178(10) & 497(5) &  & 0.2028(17) &  0.2025(22)    \\
I $^2\Delta_{5/2}$ & 2.178(10) & 497(5) &  & 0.2028(17) &  0.2026(2)  \\
\hline\hline
\end{tabular}
\end{minipage}
\end{table*}



\begin{table*}[ht]
\begin{center}
\begin{minipage}{440pt}
\caption{Composition of states in RaF in terms of Ra$^+$ valence electron configurations. Only configurations that contribute $\geq10\%$ are shown.}\label{tab:config}%
\begin{tabular}{cc}
\hline \hline
State & Composition \\ \hline
$X$~$^2 \Sigma_{1/2}$   &  90$\%$ $7s$  \\
$A$~$^2 \Pi_{1/2}$     &  60$\%$ $7p_{1/2}$ + 20$\%$ $6d_{3/2}$ + 10$\%$ $7p_{3/2}$ \\
$B$~$^2 \Delta_{3/2}$ &  40$\%$ $6d_{3/2}$ + 30$\%$ $6d_{5/2}$ + 20$\%$ $7p_{3/2}$ \\
$B$~$^2 \Delta_{5/2}$ &  90$\%$ $6d_{5/2}$ + 10$\%$ $7d_{5/2}$ \\
$A$~$^2 \Pi_{3/2}$    &  50$\%$ $7p_{3/2}$ + 40$\%$ $6d_{3/2}$ \\
$C$~$^2 \Sigma_{1/2}$ &  50$\%$ $7p_{3/2}$ + 30$\%$ $6d_{5/2}$ + 10$\%$ $7d_{5/2}$\\
{$D$~$^2 \Pi_{1/2}$} & 40$\%$ $6d_{3/2}$ + 20$\%$ $7p_{1/2}$ + 20$\%$ $6d_{5/2}$ + 10$\%$ $7p_{1/2}$ + 10$\%$ $8p_{1/2}$ \\
{$D$~$^2 \Pi_{3/2}$} & 50$\%$ $6d_{5/2}$ + 30$\%$ $7p_{3/2}$ + 10$\%$ $8p_{3/2}$ + 10$\%$ $6d_{3/2}$ + 10$\%$ $7d_{5/2}$ \\
$E$~$^2 \Sigma_{1/2}$ &  70$\%$ $8s$ + 10$\%$ $8p_{1/2}$ + 10$\%$ $9s$ \\
{$F$~$^2 \Sigma_{1/2}$} & 30$\%$ $8p_{1/2}$ + 20$\%$ $8p_{3/2}$ + 10$\%$ $7p_{3/2}$ + 10$\%$ $6d_{3/2}$ + 10$\%$ $6d_{5/2}$ +10$\%$ $8s$ \\
$G$~$^2 \Pi_{1/2}$    & 30$\%$ $8p_{3/2}$ + 30$\%$ $7d_{3/2}$ + 20$\%$ $8p_{1/2}$ + 10$\%$ $7p_{1/2}$  \\
$G$~$^2 \Pi_{3/2}$    &  50$\%$ $8p_{3/2}$ + 20$\%$ $7d_{5/2}$ + 10$\%$ $7p_{3/2}$ + 10 $\%$ $7d_{3/2}$ \\
$H$~$^2 \Sigma_{1/2}$ & 40$\%$ $7d_{5/2}$ + 10$\%$ $7p_{3/2}$ + 10$\%$ $8p_{3/2}$ + 10$\%$ $7d_{3/2}$ \\
$I$~$^2 \Delta_{3/2}$ & 60$\%$ $7p_{3/2}$ + 20$\%$ $7d_{5/2}$ \\
$I$~$^2 \Delta_{5/2}$ & 70$\%$ $7d_{5/2}$ + 10$\%$ $6d_{5/2}$\\
\hline \hline
\end{tabular}
\end{minipage}
\end{center}
\end{table*}

\begin{table*}[ht]
\begin{center}
\begin{minipage}{440pt}
\caption{Observed spin-orbit (SO) interaction constants (in cm$^{-1}$) for the states in RaF assigned in this work ($A_{\rm{RaF}}$), and states in Ra$^+$ ($\zeta_{\rm{Ra^+}}$) whose valence configuration is most dominant in the composition of the corresponding RaF state.}\label{tab:SO}%
\begin{tabular}{lcc}
\hline\hline
         & $A_{\rm{RaF}}$ & $\zeta_{\rm{Ra^+}}$\\
        \hline
$A$ $^2\Pi$  / $7p\pi$ \hspace{0.75cm}    & 2051(1) & 3238\\
$B$ $^2\Delta$ / $6d\delta$ & 404(1)  & 663 \\
$D$ $^2\Pi$ /  $6d\pi$ & 362(1)  & 663\\
$G$ $^2\Pi$ / $8p\pi$   & 452(1) & 1190\\
$I$ $^2\Delta$  /   $7d\delta$ & 54(1)  & 198\\
\hline\hline
\end{tabular}
\end{minipage}
\end{center}
\end{table*}

%